\documentclass[journal]{IEEEtran}
\usepackage{graphicx}
\usepackage{balance}
\usepackage{graphicx}
\usepackage{cite}
\usepackage{amssymb}
\usepackage{multicol}

\ifCLASSINFOpdf
\else
\fi

\begin{document}

\title{Dynamic Output Feedback Guaranteed-Cost Synchronization for Multiagent Networks with Given Cost Budgets}

\author{Jianxiang~Xi,~Cheng~Wang,~Hao~Liu,~Zhong~Wang\vspace{-1em}
\thanks{This work was supported by the National Natural
Science Foundation of China under Grants 61374054, 61503012, 61703411, 61503009, 61333011 and 61421063, Innovation Foundation of High-Tech Institute of Xi'an (2015ZZDJJ03) and Youth Foundation of High-Tech Institute of Xi'an (2016QNJJ004), also supported by Innovation Zone Project under Grants 17-163-11-ZT-004-017-01.}
\thanks{Jianxiang~Xi,~Cheng~Wang,~Zhong~Wang are with High-Tech Institute of Xi'an, Xi'an, 710025, P.R. China, and Hao~Liu is with School of Astronautics, Beihang University, Beijing, 100191, P.R. China. (Corresponding author: Hao~Liu, liuhao13@buaa.edu.cn)}}

\markboth{IEEE Access}
{Xi \MakeLowercase{\textit{et al.}}: Dynamic Output Feedback Guaranteed-Cost Synchronization for Multiagent Networks with Given Cost Budgets}

\maketitle

\begin{abstract}
The current paper addresses the distributed guaranteed-cost synchronization problems for general high-order linear multiagent networks. Existing works on the guaranteed-cost synchronization usually require all state information of neighboring agents and cannot give the cost budget previously. For both leaderless and leader-following interaction topologies, the current paper firstly proposes a dynamic output feedback synchronization protocol with guaranteed-cost constraints, which can realize the tradeoff design between the energy consumption and the synchronization regulation performance with the given cost budget. Then, according to different structure features of interaction topologies, leaderless and leader-following guaranteed-cost synchronization analysis and design criteria are presented, respectively, and an algorithm is proposed to deal with the impacts of nonlinear terms by using both synchronization analysis and design criteria. Especially, an explicit expression of the synchronization function is shown for leaderless cases, which is independent of protocol states and the given cost budget. Finally, numerical examples are presented to demonstrate theoretical results.
\end{abstract}


\IEEEpeerreviewmaketitle

\section{Introduction}\label{section1}
\IEEEPARstart{I}{n} recent years, synchronization of multiagent networks with distributed control protocols has obtained great attention by researchers from different fields, formation and containment control, sensor networks, multiple agent supporting systems, distributed computation, multiple robot systems and network congestion alleviation, {\it et al.} \cite{LiuX2017IJRNC}-\hspace{-0.5pt}\cite{Yang2017Comple}.  According to different structures, multiagent networks are usually categorized into two types: leader-following ones and leaderless ones, which are associated with leader-following synchronization and leaderless synchronization, respectively. Moreover, the motions of multiagent networks contain two parts: the whole motion and the relative motions among agents. For leader-following multiagent networks, the whole motion is the motion of the leader. However, for leaderless multiagent networks, the whole motion is associated with the interaction topology and initial states of all agents and is often described by the synchronization function. In \cite{Kaviarasan2016COMPLEXITY}, some novel conclusions for robust synchronization were given.  Sakthivel {\it et al.} \cite{Sakthivel2017TSMC} proposed an inspirational method to deal with  stochastic faulty actuator-based reliable synchronization problems. The literatures  \cite{Wen2015TNNLS}-\hspace{-0.5pt}\cite{Liu2017TSMCS} also proposed some new results on synchronization. It should be pointed out that the performance optimization was not considered in \cite{Kaviarasan2016COMPLEXITY}-\hspace{-0.5pt}\cite{Liu2017TSMCS}.

However, in practical multiagent networks, the control energy is usually limited, so it is required to simultaneously consider the following two factors: the synchronization regulation performance and the energy consumption, which can be modeled as certain optimal or suboptimal problems with different cost functions to realize the tradeoff design between them. By optimizing the cost function of each agent, some synchronization control strategies were shown to achieve global goals in \cite{Semsar-Kazerooni2009IJC} and \cite{Vamvoudakis2012Automatica}. By constructing the global performance index based on the linear quadratic cost function, Cao and Ren \cite{CaoY2010TSMC} presented an optimal synchronization criteria for first-order linear multiagent networks under the condition that the interaction topology is a complete graph. For first-order nonlinear multiagent networks, optimal synchronization criteria were proposed by convex and coercive properties of the cost function in \cite{QiuZ2016Automatica} and \cite{Xie2017SCL}. For second-order linear multiagent networks, synchronization regulation performance problems were discussed by hybrid impulsive control approaches in \cite{GuanZ2014Auto} and \cite{Hu2016JFI}, where the energy consumption was not considered. Cheng {\it et al.} \cite{ChengY2013ACC} dealt with leader-following guaranteed-cost synchronization of second-order multiagent networks, which can realize the suboptimal synchronization tracking, and investigated the applications of theoretical results to interconnected pendulums. In \cite{Semsar-Kazerooni2009IJC}-\hspace{-0.5pt}\cite{ChengY2013ACC}, the dynamics of each agent has a specific structure, which can simplify the synchronization analysis and design problems.\par

Due to the complex structure of general high-order multiagent networks, optimal synchronization is usually difficult to be achieved and guaranteed-cost synchronization is more challenging than first-order and second-order multiagent networks. Zhao {\it et al.} \cite{ZhaoY2015ISA} discussed guaranteed-cost synchronization for general high-order linear multiagent networks with the linear quadratic cost function based on state errors among neighboring agents and control inputs of all agents. Zhou {\it et al.} \cite{ZhouX2015JFI} proposed an event-triggered guaranteed-cost control method to decrease the energy consumption. In \cite{Zhao2017ISA}, sampled-data information was used to design guaranteed-cost synchronization prototols and an input delay approch was applied to give guaranteed-cost synchronization criteria. In \cite{ZhaoY2015ISA}-\hspace{-0.5pt}\cite{Zhao2017ISA}, the linear matrix inequality (LMI) synchronization design criteria contain the Laplacian matrix and the dimensions of variables are associated with the number of agents, which cannot ensure the scalability of multiagent networks since the computational complexity greatly increases as the number of agents increases. To overcome this flaw, the state decomposition approach was shown to deal with guaranteed-cost synchronization in \cite{XiJ2016JFI}-\hspace{-0.5pt}\cite{XiJ2018IJRCN}, where LMI synchronization design criteria are only dependent on the nonzero eigenvalues of the Laplacian matrix and the dimensions of all the variables are identical with the one of each agent. Moreover, Xie and Yang \cite{Xie2016Neurocomputing} proposed sufficient conditions for guaranteed-cost fault-tolarant synchronization by introducing a coupling weight larger than the reciprocal of the minimum nonzero eigenvalue of the Laplacian matrix, where the dimension of the variable of the algebraic Riccati equality is independent of the number of agents.\par

Although some significant research results on guaranteed-cost synchronization were presented, there still exist many very challenging and open problems. The current paper mainly focuses on the following two aspects: (i) The cost budget is given previously. For practical multiagent networks, each agent usually has the limited energy, so the cost budget cannot be infinite and should be a finite value given previously. In \cite{ZhaoY2015ISA}-\hspace{-0.5pt}\cite{Xie2016Neurocomputing}, different upper bounds of the guaranteed cost were determined, but they cannot be given previously; (ii) The outputs instead of the states of neighboring agents are used to construct the synchronization protocol. In practical applications, each agent often can only observe its neighbors and obtain output information which may be partial states or linear combinations of states. It is well-known that output feedback synchronization control is more complex and challengeable than state feedback synchronization control. In \cite{ZhaoY2015ISA}-\hspace{-0.5pt}\cite{Xie2016Neurocomputing}, all state information of neighboring agents is required to realize the guaranteed-cost synchronization control.

For leaderless and leader-following general high-order linear multiagent networks with the given cost budgets, the current paper proposes a dynamic output feedback synchronization protocol with a specific structure to deal with guaranteed-cost synchronization analysis and design problems. For leaderless cases, the relationship between the given cost budget and the LMI variable is constructed by initial states of all agents and the Laplacian matrix of a complete graph, guaranteed-cost synchronization analysis and design criteria are proposed, respectively, and the synchronization function is determined. For leader-following cases, the relationship between the given cost budget and the LMI variable is determined via initial states of all agents and the Laplacian matrix of a star graph, and sufficient conditions for guaranteed-cost synchronization criteria are presented by LMI tools. Moreover, based on the cone complementarity approach, an algorithm is proposed to check guaranteed-cost synchronization design criteria which contain nonlinear matrix inequality constraints.

Compared with closely related works on guaranteed-cost synchronization, the current paper has two critical innovations. The first one is that the cost budget is given previously in the current paper. The literatures \cite{ZhaoY2015ISA}-\hspace{-0.5pt}\cite{Xie2016Neurocomputing} only determined different upper bounds of the guaranteed cost, but cannot previously give the cost budget. The second one is that the current paper proposes dynamic output feedback synchronization protocols with the linear quadratic optimization index. The literatures \cite{ZhaoY2015ISA}-\hspace{-0.5pt}\cite{Xie2016Neurocomputing} required all state information of neighboring agents to construct guaranteed-cost synchronization protocols.

The remainder of the current paper is organized as follows. In Section \ref{section2}, some preliminaries on graph theory and the problem description are presented, respectively. Section \ref{section3} gives guaranteed-cost synchronization criteria for leaderless multiagent networks with dynamic output feedback synchronization protocols and the given cost budget, and determines an explicit expression of the synchronization function. Section \ref{section4} presents leader-following guaranteed-cost synchronization criteria. Section \ref{section5} shows numerical examples to illustrate theoretical results. Some concluding remarks are given in Section \ref{section6}.

{\it Notations}: ${{\bf R}^n}$ is the  $n$-dimensional real column vector space and ${{\bf R}^{n \times n}}$ is the set of ${n \times n}$ dimensional real matrices. ${I_n}$ represents the $n$-dimensional identity matrix. ${\bf{1}}$ denotes a column vector with all components 1. 0 and ${\bf{0}}$ stand for the zero number and the zero column vector with a compatible dimension, respectively. The notation $*$ in a symmetric matrix denotes the symmetric term. The symbol $ \otimes $ represents the Kronecker product. ${P^T} = P < {\rm{0}}$ and ${P^T} = P > {\rm{0}}$ mean that the symmetric matrix $P$ is negative definite and positive definite, respectively. The notation ${\rm{diag}}\left\{ {{d_1},{d_2}, \cdots ,{d_N}} \right\}$ represents a diagonal matrix with the diagonal elements ${d_1},{d_2}, \cdots ,{d_N}$. The notation ${\rm{tr}}\left( P \right)$ denotes the trace of the matrix $P$.
\par \vspace{-0.25em}

\section{Preliminaries and problem description}\label{section2}

\subsection{Preliminaries on graph theory}
The current paper models the interaction topology of a multiagent network with $N$ identical agents by a graph $G{\rm{ = }}\left( {V(G),E(G)} \right)$, which is composed by a nonempty vertex set $V(G) = \left\{ {{v_1},{v_2}, \cdots ,{v_N}} \right\}$ and the edge set $E(G) = \left\{ {{e_{ij}} = ({v_i},{v_j})} \right\}$. The vertex ${v_i}$ represents agent $i$, the edge ${e_{ij}}$ denotes the interaction channel from agent $i$ to agent $j$, and the edge weight ${w_{ji}}$ of ${e_{ij}}$ stands for the interaction strength from agent $i$ to agent $j$. The index of the set of all neighbors of vertex ${v_j}$ is denoted by ${N_j} = \left\{ {i:({v_i},{v_j}) \in E(G)} \right\}$. A path between vertex ${v_{{i_1}}}$ and vertex ${v_{{i_l}}}$ is a sequence of edges $\left( {{v_{{i_1}}},{v_{{i_2}}}} \right),\left( {{v_{{i_2}}},{v_{{i_3}}}} \right), \cdots ,\left( {{v_{{i_{l - 1}}}},{v_{{i_l}}}} \right)$. An undirected graph is said to be connected if there at least exists an undirected path between any two vertices. A directed graph has a spanning tree if there exists a root node which has a directed path to any other nodes. Define the Laplacian matrix of the graph $G$ as   $L = \left[ {{l_{ji}}} \right] \in {{\bf R}^{N \times N}}$ with ${l_{jj}} = \sum\nolimits_{i \in {N_j}} {{w_{ji}}}$ and ${l_{ji}} =  - {w_{ji}}{\rm{ }}\left( {j \ne i} \right)$. If the undirected graph is connected, then zero is a simple eigenvalue of $L$, and all the other $N-1$ eigenvalues are positive. If the directed graph has a spanning tree, then zero is a simple eigenvalue of $L$, and all the other $N-1$ eigenvalues have positive real parts. More basic concepts and conclusions on graph theory can be found in \cite{GodsilC2001}. \vspace{-0.25em}

\subsection{Problem description}
For multiagent networks consisting of $N$ identical high-order linear agents, the dynamics of the $j$th agent is described by
\begin{eqnarray}\label{1}
\left\{ \begin{array}{l}
 {{\dot x}_j}(t) = A{x_j}(t) + B{u_j}(t), \\
 {y_j}(t) = C{x_j}(t), \\
 \end{array} \right.
\end{eqnarray}
where $j = 1,2, \cdots ,N$, $A \in {{\bf R} ^{n \times n}}$, $B \in {{\bf R} ^{n \times m}}$, $C \in {{\bf R} ^{d \times n}}$ and ${x_j}(t)$, ${y_j}(t)$ and ${u_j}(t)$ are the state, the output and the control input, respectively. For ${Q^T} = Q > 0$ and ${R^T} = R > 0$, a dynamic output feedback synchronization protocol with a linear quadratic optimization index is proposed as follows:
\begin{eqnarray}\label{2}
\left\{ \begin{array}{l}
 {{\dot \phi }_j}(t) = \left( {A + B{K_u}} \right){\phi _j}(t) \\
  {\kern 32pt}- {K_\phi }C\sum\limits_{i \in {N_j}} {{w_{ji}}\left( {{\phi _i}(t) - {\phi _j}(t)} \right)} \\
  {\kern 32pt}+ {K_\phi }\sum\limits_{i \in {N_j}} {{w_{ji}}\left( {{y_i}(t) - {y_j}(t)} \right),}  \\
 {u_j}(t) = {K_u}{\phi _j}(t), \\
 {J_{\rm{s}}} = \int_0^\infty  {\left( {{J_u}(t) + {J_{x\phi }}(t)} \right){\rm{d}}t} , \\
 \end{array} \right.
\end{eqnarray}
where $j = 1,2, \cdots ,N$, ${\phi _j}(t)$ with ${\phi _j}(0) = {\bf{0}}$ is the protocol state, ${K_u}$ and ${K_\phi }$ are gain matrices with compatible dimensions to be determined, ${N_j}$ represents the neighbor set of agent ${j}$ and
\[
{J_u}(t) = \sum\limits_{j = 1}^N {u_j^T(t)R{u_j}(t)}, \]
\[
 {J_{x\phi }}(t) = \sum\limits_{j = 1}^N {\sum\limits_{i \in {N_j}} {\left( {{w_{ji}}{{\left( {{x_i}(t) - {x_j}(t) - {\phi _i}(t) + {\phi _j}(t)} \right)}^T}} \right.} }
\]\vspace{-10pt}
\[
 \hspace{72pt}\left. { \times Q\left( {{x_i}(t) - {x_j}(t) - {\phi _i}(t) + {\phi _j}(t)} \right)} \right) .
\]

Furthermore, ${J_u}(t)$ and ${J_{x\phi }}(t)$ are called the energy consumption term and the synchronization regulation term, respectively, and the tradeoff design between the energy consumption and the synchronization regulation performance can be realized by choosing proper $R$ and $Q$. It should be pointed out that there also exists the linear quadratic index to realize guaranteed-cost control for isolated systems as shown in \cite{Thuan2012AMC}, but its structure is different with the one in (\ref{2}). For isolated systems, the linear quadratic index is constructed by state information, which is convergent. For multiagent networks, it is required that state errors among agents are convergent, but states of each agent may be divergent. Hence, the linear quadratic index for multiagent networks should be constructed by state errors as shown in (\ref{2}), and cannot use state information. Furthermore, guaranteed-cost control can be clarified into two types. The first one is to calculate the upper bound of the linear quadratic index for given gain matrices as shown in \cite{ZhaoY2015ISA}-\hspace{-0.5pt}\cite{Xie2016Neurocomputing}. The second one is to determine gain matrices of synchronization protocols for the given upper bound of the linear quadratic index; that is, the given cost budget. Moreover, it can be shown that $- {K_\phi }C\sum\nolimits_{i \in {N_j}} {{w_{ji}}\left( {{\phi _i}(t) - {\phi _j}(t)} \right)}  + {K_\phi }\sum\nolimits_{i \in {N_j}} {{w_{ji}}\left( {{y_i}(t) - {y_j}(t)} \right)}  = {K_\phi }C\sum\nolimits_{i \in {N_j}} {{w_{ji}}\left( {{x_i}(t)} \right.}  - {x_j}(t) \left. { - {\phi _i}(t) + {\phi _j}(t)} \right) $,
which means that the term $\sum\nolimits_{i \in {N_j}} {{w_{ji}}} \left( {{x_i}(t) - } \right.$ $\left. {{x_j}(t) - {\phi _i}(t) + {\phi _j}(t)} \right)$ directly impacts on the derivative of the protocol state and indirectly impacts on the derivative of the state of each agent. Hence, we choose ${J_{x\phi }}(t)$ as the index function of the synchronization regulation performance.

Let $J_{\rm{s}}^* > 0$ be a given cost budget, then the definition of guaranteed-cost synchronization of multiagent networks with the given cost budget is proposed as follows.

{\emph{Definition 1:}}~~For any given $J_{\rm{s}}^* > 0$, multiagent network (\ref{1}) is said to be {\it guaranteed-cost synchronizable} by protocol (\ref{2}) if there exist ${K_u}$ and ${K_\phi }$ such that ${\lim _{t \to \infty }}\left( {{x_j}(t) - c(t)} \right) = {\bf{0}}$ $\left( {j = 1,2, \cdots ,N} \right)$ and ${J_{\rm{s}}} \le J_{\rm{s}}^*$ for any bounded disagreement initial states ${x_j}(0){\rm{ }}(j = 1,2, \cdots ,N)$, where $c(t)$ is said to be the {\it synchronization function}.

The main objects of the current paper are to design ${K_u}$ and ${K_\phi }$ such that multiagent network (\ref{1}) with leaderless and leader-following structures achieves guaranteed-cost synchronization under the condition that the cost budget is given, and to determine the impacts of the state of the synchronization protocol and the given cost budget on the synchronization function for leaderless cases.

{\emph{Remark 1:}} Compared with guaranteed-cost synchronization protocols in \cite{ZhaoY2015ISA}-\hspace{-0.5pt}\cite{Xie2016Neurocomputing}, protocol (\ref{2}) has two critical features. The first one is that outputs instead of states of neighboring agents are applied to construct synchronization protocols. For dynamic output feedback synchronization protocols, the key challenge is that the upper bound of the optimization index is difficult to be determined since both the energy consumption term and the synchronization regulation term are dependent on protocol states. The second one is that the cost budget is given previously. In this case, the key challenge is to determine the relationship between the upper bound of the optimization index and the given cost budget and to design gain matrices of synchronization protocols such that the upper bound is less than the given cost budget. Moreover, compared with the traditional dynamic output feedback controller for isolated systems as shown in classic literatures \cite{Thuan2012IJAMCS} and \cite{Thuan2012AMC}, the key difference is that output errors between one agent and its neighbors are used to construct synchronization protocols for multiagent networks as shown in (\ref{2}). It should be pointed out that the state of each agent may be not convergent, but it is required that state errors among all agents are convergent under protocol (\ref{2}). However, it is needed that the states of an isolated system are convergent by designing the dynamic output feedback controller.

\section{Guaranteed-cost synchronization for leaderless multiagent networks}\label{section3}
For high-order linear multiagent networks with leaderless connected topologies, this section gives sufficient conditions for guaranteed-cost synchronization design and analysis with the given cost budget, respectively, where the guaranteed-cost synchronization design criterion contains a nonlinear constraint, so an algorithm is proposed to determine gain matrices on the basis of the cone complementarity approach. Moreover, an explicit expression of the synchronization function is shown, which is independent of the protocol state and the given cost budget.

Let $x(t) = {\left[ {x_1^T(t),x_2^T(t), \cdots ,x_N^T(t)} \right]^T}$ and $\phi (t) = {[ {\phi _1^T(t),}}$ ${{\phi _2^T(t), \cdots ,\phi _N^T(t)} ]^T}{\rm{,}}$ then the dynamics of multiagent network (\ref{1}) with protocol (\ref{2}) can be written as
\begin{eqnarray}\label{3}
\left\{ \begin{array}{l}
 \dot x(t) = \left( {{I_N} \otimes A} \right)x(t) + \left( {{I_N} \otimes B{K_u}} \right)\phi (t), \\
 \dot \phi (t) = \left( {{I_N} \otimes \left( {A + B{K_u}} \right) + \left( {L \otimes {K_\phi }C} \right)} \right)\phi (t) \\
  {\kern 28pt}  - \left( {L \otimes {K_\phi }C} \right)x(t). \\
 \end{array} \right.
\end{eqnarray}
Because the interaction topology is undirected, the Laplacian matrix $L$ is symmetric and positive semi-definite. Due to $L{\bf{1}} = {\bf{0}}$, there exists an orthonormal matrix $U = \left[ {{{\bf{1}} \mathord{\left/ {\vphantom {{\bf{1}} {\sqrt N }}} \right.  \kern-\nulldelimiterspace} {\sqrt N }},\hat U} \right]$ such that ${U^T}LU = {\rm{diag}}\left\{ {0,\Delta } \right\}$, where $\Delta  = {\rm{diag}}\left\{ {{\lambda _2},{\lambda _3}, \cdots ,{\lambda _N}} \right\}$ with $0 < {\lambda _2} \le {\lambda _3} \le  \cdots  \le {\lambda _N}$. Let
\begin{eqnarray}\label{4}
\hat x(t) = \left( {{U^T} \otimes {I_n}} \right)x(t) = {\left[ {\hat x_1^T(t),\hat x_2^T(t), \cdots ,\hat x_N^T(t)} \right]^T}{\rm{,}}
\end{eqnarray}
\begin{eqnarray}\label{5}
\hat \phi (t) = \left( {{U^T} \otimes {I_n}} \right)\phi (t) = {\left[ {\hat \phi _1^T(t),\hat \phi _2^T(t), \cdots ,\hat \phi _N^T(t)} \right]^T},
\end{eqnarray}
then multiagent network (\ref{3}) can be transformed into
\begin{eqnarray}\label{6}
\left\{ {\begin{array}{*{20}{c}}
   {{{\dot {\hat x}}_1}(t) = A{{\hat x}_1}(t) + B{K_u}{{\hat \phi }_1}(t),} \hfill  \\
   {{{\dot {\hat \phi }}_1}(t) = \left( {A + B{K_u}} \right){{\hat \phi }_1}(t),} \hfill  \\
\end{array}} \right.
\end{eqnarray}
\begin{eqnarray}\label{7}
\left\{ \begin{array}{l}
 {{\dot {\hat x}}_j}(t) = A{{\hat x}_j}(t) + B{K_u}{{\hat \phi }_j}(t), \\
 {{\dot {\hat \phi} }_j}(t) = \left( {A + B{K_u} + {\lambda _j}{K_\phi }C} \right){{\hat \phi }_j}(t) \\
   {\kern 33pt}  - {\lambda _j}{K_\phi }C{{\hat x}_j}(t), \\
 \end{array} \right.
\end{eqnarray}
where $j = 2,3, \cdots ,N$.

The $N$-dimensional column vector with the $j$th element 1 and 0 elsewhere is denoted by ${e_j}$ $(j = 2,3, \cdots ,N)$. Define
\begin{eqnarray}\label{8}
{x_{\rm{e}}}(t) \buildrel \Delta \over = \sum\limits_{j = 2}^N {U{e_j} \otimes {{\hat x}_j}(t)},
\end{eqnarray}
\begin{eqnarray}\label{9}
{x_{\rm{s}}}(t) \buildrel \Delta \over = \frac{1}{{\sqrt N }}{\bf{1}} \otimes {\hat x_1}(t),
\end{eqnarray}
then one can show by (\ref{8}) that
\begin{eqnarray}\label{10}
{x_{\rm{e}}}(t) = \left( {U \otimes {I_n}} \right){\left[ {{{\bf{0}}^T},\hat x_2^T(t),\hat x_3^T(t), \cdots ,\hat x_N^T(t)} \right]^T}.
\end{eqnarray}
By $U{e_1} = {{\bf{1}} \mathord{\left/
 {\vphantom {{\bf{1}} {\sqrt N }}} \right.
 \kern-\nulldelimiterspace} {\sqrt N }}$ and ${e_1} \otimes {\hat x_1}(t) = {\left[ {\hat x_1^T(t),{{\bf{0}}^T}} \right]^T}$, it can be derived from (\ref{9}) that
\begin{eqnarray}\label{11}
{x_{\rm{s}}}(t) = \left( {U \otimes {I_n}} \right){\left[ {\hat x_1^T(t),{{\bf{0}}^T}} \right]^T}.
\end{eqnarray}
Since $U$ is nonsingular, ${x_{\rm{e}}}(t)$ and ${x_{\rm{s}}}(t)$ are linearly independent by (\ref{10}) and (\ref{11}). From (\ref{4}), one can obtain that $x(t) = {x_{\rm{e}}}(t) + {x_{\rm{s}}}(t)$. By the structure of ${x_{\rm{s}}}(t)$ given in (\ref{9}), multiagent network (\ref{3}) achieves leaderless synchronization if and only if ${\lim _{t \to \infty }}{\hat x_j}(t) = {\bf{0}}{\rm{ }}\left( {j = 2,3, \cdots ,N} \right)$ and ${{{{\hat x}_1}(t)} \mathord{\left/ {\vphantom {{{{\hat x}_1}(t)} {\sqrt N }}} \right.  \kern-\nulldelimiterspace} {\sqrt N }}$ is a valid candidate of the synchronization function. Thus, ${x_{\rm{e}}}(t)$ and ${x_{\rm{s}}}(t)$ can be regarded as the error state among agents and the synchronization state of multiagent network (\ref{3}), which stands for the disagreement part and the agreement part, respectively. Furthermore, one can find by (\ref{7}) that ${\lim _{t \to \infty }}[\hat \phi _j^T(t),$ $\hat x_j^T(t){]^T} = {\bf{0}}$ $\left( {j = 2,3, \cdots ,N} \right)$ can guarantee that multiagent network (\ref{1}) with protocol (\ref{2}) achieves leaderless synchronization.

Based on the above analysis, the following theorem presents an approach to determine gain matrices ${K_u}$ and ${K_\phi }$ such that multiagent network (\ref{1}) with protocol (\ref{2}) achieves leaderless guaranteed-cost synchronization with a given cost budget.

{\emph{Theorem 1:}}~~For any given $J_{\rm{s}}^* > 0$, multiagent network (\ref{1}) is leaderless guaranteed-cost synchronizable by protocol (\ref{2}) if there exist $P_x^T = {P_x} > 0,$ $\hat P_x^T = {\hat P_x} > 0,$ $\hat P_\phi ^T = {\hat P_\phi } > 0,$ and ${\hat K_u }$ such that
\[
{\hat \Xi _1} = {x^T}(0)\left( {\left( {{I_N} - {N^{ - 1}}{\bf{1}}{{\bf{1}}^T}} \right) \otimes {I_n}} \right)x(0){P_x} - J_{\rm{s}}^*{I_n} \le 0,\]
\[
{\hat \Xi _j} = \left[ {\begin{array}{*{20}{c}}
   {{\Xi _{11}}} & { - {\lambda _j}{{\hat P}_x}{C^T}C} & {\hat K_u^TR}  \\
   {*} & {\Xi _{22}^j} & {\bf{0}}  \\
   {*} & {\bf{0}} & { - R}  \\
\end{array}} \right] < 0\left( {j = 2,N} \right),\]
\[
{P_x}{\hat P_x} = {I_n},\]
where ${\Xi _{11}} = A{\hat P_\phi } + {\hat P_\phi }{A^T} + B{\hat K_u} + \hat K_u^T{B^T}$ and $\Xi _{22}^j = {P_x}A + {A^T}{P_x} - 2{\lambda _j}{C^T}C + 2{\lambda _j}Q.$ In this case, ${K_u} = {\hat K_u}\hat P_\phi ^{ - 1}$ and ${K_\phi } =  - {\hat P_x}{C^T}$.

{\emph{Proof:}}~~ First of all, we give sufficient conditions by LMI techniques such that ${\lim _{t \to \infty }}{\left[ {\hat \phi _j^T(t),\hat x_j^T(t)} \right]^T} = {\bf{0}}$ $( {j =}$${2,3,\cdots ,N} )$. One can derive that
\begin{eqnarray}\label{13}
\left[ {\begin{array}{*{20}{c}}
   {{{\hat \phi }_j}(t)}  \\
   {{{\hat \phi }_j}(t) - {{\hat x}_j}(t)}  \\
\end{array}} \right] = \left[ {\begin{array}{*{20}{c}}
   {{I_n}} & {\bf{0}}  \\
   {{I_n}} & { - {I_n}}  \\
\end{array}} \right]\left[ {\begin{array}{*{20}{c}}
   {{{\hat \phi }_j}(t)}  \\
   {{{\hat x}_j}(t)}  \\
\end{array}} \right],
\end{eqnarray}
so subsystems (\ref{7}) can be converted into
\begin{eqnarray}\nonumber
\begin{array}{l}
 \left[ {\begin{array}{*{20}{c}}
   {{{\dot {\hat \phi} }_j}(t)}  \\
   {{{\dot {\hat \phi} }_j}(t) - {{\dot {\hat x}}_j}(t)}  \\
\end{array}} \right] = \left[ {\begin{array}{*{20}{c}}
   {A + B{K_u}} & {{\lambda _j}{K_\phi }C}  \\
   {\bf{0}} & {A + {\lambda _j}{K_\phi }C}  \\
\end{array}} \right]
 \end{array}
 \end{eqnarray}\vspace{-8pt}
\begin{eqnarray}\label{14}
  {\kern 60pt}  \times \left[ {\begin{array}{*{20}{c}}
   {{{\hat \phi }_j}(t)}  \\
   {{{\hat \phi }_j}(t) - {{\hat x}_j}(t)}
\end{array}} \right].
 \end{eqnarray}
Let ${P_\phi }$ and ${P_x}$ be symmetric and positive definite matrices, then we construct a Lyapunov function candidate as follows
\begin{eqnarray}\label{15}
{V_j}(t) = {V_{\phi j}}(t) + {V_{xj}}(t),
\end{eqnarray}
where $j = 2,3, \cdots ,N$ and
\[
{V_{\phi j}}(t) = \hat \phi _j^T(t){P_\phi }{\hat \phi _j}(t),\]
\[
{V_{xj}}(t) = {\left( {{{\hat \phi }_j}(t) - {{\hat x}_j}(t)} \right)^T}{P_x}\left( {{{\hat \phi }_j}(t) - {{\hat x}_j}(t)} \right).\]
From (\ref{14}) to (\ref{15}), one can show that
\[
\begin{array}{l}
 {{\dot V}_{\phi j}} = \hat \phi _j^T(t)\left( {{P_\phi }\left( {A + B{K_u}} \right) + {{\left( {A + B{K_u}} \right)}^T}{P_\phi }} \right){{\hat \phi }_j}(t) \\
 {\kern 26pt} + 2{\lambda _j}\hat \phi _j^T(t){P_\phi }{K_\phi }C\left( {{{\hat \phi }_j}(t) - {{\hat x}_j}(t)} \right), \\
 \end{array}\]
\[\begin{array}{l}
 {{\dot V}_{xj}} = {\left( {{{\hat \phi }_j}(t) - {{\hat x}_j}(t)} \right)^T}\left( {{P_x}\left( {A + {\lambda _j}{K_\phi }C} \right)} \right. \\
 {\kern 25pt} \left. { + {{\left( {A + {\lambda _j}{K_\phi }C} \right)}^T}{P_x}} \right)\left( {{{\hat \phi }_j}(t) - {{\hat x}_j}(t)} \right). \\
 \end{array}\]
Thus, it can be derived that ${\lim _{t \to \infty }}{\hat \phi _j}(t) = {\bf{0}}$ and ${\lim _{t \to \infty }}\left( {{{\hat \phi }_j}(t) - {{\hat x}_j}(t)} \right) = {\bf{0}}$ if
\begin{eqnarray}\label{16}
\begin{array}{l}
 {\Theta _j} = \left[ {\begin{array}{*{20}{c}}
   {{P_\phi }\left( {A + B{K_u}} \right) + {{\left( {A + B{K_u}} \right)}^T}{P_\phi }}  \\
   {*}  \\
\end{array}} \right. \\
 {\kern 20pt} \left. {\begin{array}{*{20}{c}}
   {{\lambda _j}{P_\phi }{K_\phi }C}  \\
   {{P_x}\left( {A + {\lambda _j}{K_\phi }C} \right) + {{\left( {A + {\lambda _j}{K_\phi }C} \right)}^T}{P_x}}  \\
\end{array}} \right] < 0, \\
 \end{array}
\end{eqnarray}
where $j = 2,3, \cdots ,N$, which means that multiagent network (\ref{1}) with protocol (\ref{2}) achieves leaderless synchronization due to ${\lim _{t \to \infty }}{\left[ {\hat \phi _j^T(t),\hat x_j^T(t)} \right]^T} = {\bf{0}}$ $(j = 2,3, \cdots ,N)$.

In the following, the guaranteed-cost performance is discussed. Due to ${\phi _j}(0) = {\bf{0}}{\rm{ }}\left( {j = 1,2, \cdots ,N} \right)$, one can show that ${\hat \phi _1}(0) = {\bf{0}}$. By (\ref{6}), one has ${\hat \phi _1}(t) \equiv {\bf{0}}$. Thus, it can be obtained by (\ref{4}) and (\ref{5}) that
\begin{eqnarray}\nonumber
{J_u}(t) = {\phi ^T}(t)\left( {{I_N} \otimes K_u^TR{K_u}} \right)\phi (t)
\end{eqnarray}\vspace{-20pt}
\begin{eqnarray}\label{17}
{\kern 12pt} = \sum\limits_{j = 2}^N {\hat \phi _j^T(t)K_u^TR{K_u}{{\hat \phi }_j}(t)} ,
\end{eqnarray}
\begin{eqnarray}\nonumber
 \hspace{-2em}{J_{x\phi }}(t) = {\left( {\phi (t) - x(t)} \right)^T}\left( {2L \otimes Q} \right)\left( {\phi (t) - x(t)} \right)
\end{eqnarray}\vspace{-20pt}
\begin{eqnarray}\label{18}
{\kern 34pt} = \sum\limits_{j = 2}^N {2{\lambda _j}{{\left( {{{\hat \phi }_j}(t) - {{\hat x}_j}(t)} \right)}^T}Q\left( {{{\hat \phi }_j}(t) - {{\hat x}_j}(t)} \right)} . \end{eqnarray}
For $T \ge 0$, we can derive from (\ref{16}) to (\ref{18}) that
\[
\begin{array}{l}
 {J_{{\rm{s}}T}} \buildrel \Delta \over = \int_0^T {\left( {{J_u}(t) + {J_{x\phi }}(t)} \right){\rm{d}}t}  \\
  {\kern 15pt} = \int_0^T {\left( {{J_u}(t) + {J_{x\phi }}(t)} \right){\rm{d}}t}  \\
  {\kern 25pt} + \sum\limits_{j = 2}^N {\left( {\int_0^T {{{\dot V}_j}(t){\rm{d}}t}  - {V_j}(T) + {V_j}(0)} \right)}  \\
  {\kern 15pt}  = \sum\limits_{j = 2}^N {\int_0^T {\left( {\hat \phi _j^T(t){P_\phi }\left( {\left( {A + B{K_u}} \right)P_\phi ^{ - 1}} \right.} \right.} }  \\
  {\kern 25pt}  \left. { + P_\phi ^{ - 1}{{\left( {A + B{K_u}} \right)}^T} + P_\phi ^{ - 1}K_u^TR{K_u}P_\phi ^{ - 1}} \right){P_\phi }{{\hat \phi }_j}(t) \\
 {\kern 25pt} +  \hspace{-2pt}2{\lambda _j}\hat \phi _j^T(t){P_\phi }{K_\phi }C\left( {{{\hat \phi }_j}(t) \hspace{-2pt} - \hspace{-2pt} {{\hat x}_j}(t)} \right)  \hspace{-2pt}+ \hspace{-2pt} {\left( {{{\hat \phi }_j}(t) \hspace{-2pt} -\hspace{-2pt}  {{\hat x}_j}(t)} \right)^T} \\
 {\kern 25pt}  \times \left( {{P_x}\left( {A + {\lambda _j}{K_\phi }C} \right) + {{\left( {A + {\lambda _j}{K_\phi }C} \right)}^T}{P_x} + 2{\lambda _j}Q} \right) \\
 {\kern 25pt}  \times \left. {\left( {{{\hat \phi }_j}(t) - {{\hat x}_j}(t)} \right)} \right){\rm{d}}t  - \sum\limits_{j = 2}^N {\left( {{V_j}(T) - {V_j}(0)} \right)}.  \\
 \end{array}
 \]
Let ${\hat K_u} = {K_u}{\hat P_\phi }$ with ${\hat P_\phi } = P_\phi ^{ - 1}$ and ${K_\phi } =  - {\hat P_x}{C^T}$ with ${\hat P_x} = P_x^{ - 1}$. By Schur Complement Lemma in \cite{Boyad1994} , if ${\hat \Xi _j} < 0$ $\left( {j = 2,3, \cdots ,N} \right)$, then as $T$ tends to infinity, one has
\[
{J_{\rm{s}}} \le \sum\limits_{j = 2}^N {{V_j}(0)}.
 \]
Due to ${\phi _j}(0) = {\bf{0}}{\rm{ }}$~$\left( {j = 1,2, \cdots ,N} \right)$, one has ${\hat \phi _j}(0) = {\bf{0}}{\rm{ }}$~$\left( {j = 1,2, \cdots ,N} \right)$ by (\ref{5}), which means that ${V_{\phi j}}(0) = 0$ and ${V_{xj}}(0) = \hat x_j^T(0){P_x}{\hat x_j}(0)$. Thus, one can find that
\[
\hspace{-11em} {J_{\rm{s}}} \le \sum\limits_{j = 2}^N {\hat x_j^T(0){P_x}{{\hat x}_j}(0)}
 \]\vspace{-4pt}
\[
 {\kern 0pt} = {x^T}(0)\left( {U \otimes {I_n}} \right)\left[ {\begin{array}{*{20}{c}}
   {{{\bf{0}}^T}}  \\
   {{I_{(N - 1)n}}}  \\
\end{array}} \right]\left( {{I_{N - 1}} \otimes {P_x}} \right)
\]\vspace{-14pt}
\begin{eqnarray}\label{19}
 {\kern -12pt} \times \left[ {{\bf{0}},{I_{(N - 1)n}}} \right]\left( {{U^T} \otimes {I_n}} \right)x(0).
\end{eqnarray}
Since $U{U^T} = {I_N}$, it can be shown that
\begin{eqnarray}\label{20}
\hat U{\hat U^T} = {I_N} - {N^{ - 1}}{\bf{1}}{{\bf{1}}^T}.
\end{eqnarray}
Due to
\[
\left[ {{\bf{0}},{I_{(N - 1)n}}} \right]\left( {{U^T} \otimes {I_n}} \right) = {\hat U^T} \otimes {I_n},\]
one can derive by (\ref{19}) and (\ref{20}) that
\begin{eqnarray}\label{21}
{J_{\rm{s}}} \le {x^T}(0)\left( {\left( {{I_N} - {N^{ - 1}}{\bf{1}}{{\bf{1}}^T}} \right) \otimes {P_x}} \right)x(0).
\end{eqnarray}
Because ${x_j}(0){\rm{ }}\left( {j = 1,2, \cdots ,N} \right)$ are disagreement, there exists some ${\hat x_j}(0) \ne 0{\rm{ }}\left( {j \in \left\{ {2,3, \cdots ,N} \right\}} \right)$. Thus, one can derive that
\[
{x^T}(0)\left( {\left( {{I_N} - {N^{ - 1}}{\bf{1}}{{\bf{1}}^T}} \right) \otimes {I_n}} \right)x(0) = \sum\limits_{j = 2}^N {\hat x_j^T(0){{\hat x}_j}(0)}  > 0.\]
Hence, one can set that
\[
\gamma  = \frac{{J_{\rm{s}}^*}}{{{x^T}(0)\left( {\left( {{I_N} - {N^{ - 1}}{\bf{1}}{{\bf{1}}^T}} \right) \otimes {I_n}} \right)x(0)}};\]
that is,
\begin{eqnarray}\label{22}
J_{\rm{s}}^* = {x^T}(0)\left( {\left( {{I_N} - {N^{ - 1}}{\bf{1}}{{\bf{1}}^T}} \right) \otimes \gamma {I_n}} \right)x(0).
\end{eqnarray}
Since ${I_N} - {N^{ - 1}}{\bf{1}}{{\bf{1}}^T}$ has a simple zero eigenvalue and $N-1$ nonzero eigenvalues, ${P_x} \le \gamma {I_n}$ can guarantee that ${J_{\rm{s}}} \le J_{\rm{s}}^*$ by (\ref{21}) and (\ref{22}). Based on the above analysis, by the convex property of LMIs, the conclusion of Theorem 1 can be obtained. $\Box$

{\emph{Remark 2:}} The specific structures of coefficient matrices of protocol (\ref{2}) make subsystems (\ref{7}) satisfy some separation principle; that is, their dynamics can transformed into the ones in (\ref{14}). In this case, ${K_u}$ and ${K_\phi }$ can be independently designed such that $A + B{K_u}$ and $A + {\lambda _j}{K_\phi }C$ $\left( {j = 2,3, \cdots ,N} \right)$ are Hurwitz, which can guarantee that multiagent network (\ref{1}) with protocol (\ref{2}) but without the optimization index ${J_{\rm{s}}}$ achieves leaderless synchronization. However, when the guaranteed-cost performance is considered, the impacts of the term ${\lambda _j}{K_\phi }C$ in (\ref{14}) cannot be neglected since ${\hat \phi _j}(t) - {\hat x_j}(t)$ can directly influence the derivative of ${\hat \phi _j}(t)$ via the term ${\lambda _j}{K_\phi }C$. In this case, by left- and right-multiplying ${\Theta _j}$ $\left( {j = 2,3, \cdots ,N} \right)$ with ${\rm{diag}}\left\{ {P_\phi ^{ - 1},{I_n}} \right\}$, ${K_u}$ can be determined but ${K_\phi }$ cannot. Here, by introducing a specific structure ${K_\phi } =  - {\hat P_x}{C^T}$, the gain matrices ${K_u}$ and ${K_\phi }$ can be determined simultaneously.

{\emph{Remark 3:}} In the associated works about guaranteed-cost control, the value of the Lyapunov function candidate at time zero is used to determine the guaranteed cost. Since ${\hat \phi _j}(t)$ and ${\hat x_j}(t)$ in (\ref{7}) couple with each other, it seems difficult to construct a Lyapunov function candidate such that the expression of the upper bound of ${J_{\rm{s}}}$ does not contain initial states of synchronization protocols. Based on the separation principle, a Lyapunov function candidate is proposed in (\ref{15}), which makes an upper bound of ${J_{\rm{s}}}$ only dependent on initial states of all agents under the assumption that initial states of protocol (\ref{2}) are zero. In this case, the relationship between the upper bound of ${J_{\rm{s}}}$ and $J_{\rm{s}}^*$ can be determined by the property of ${I_N} - {N^{ - 1}}{\bf{1}}{{\bf{1}}^T}$, which actually is the Laplacian matrix of a complete graph with edge weights equal to ${N^{ - 1}}$. It should be pointed out that it will become very difficult to determine the relationship between ${J_{\rm{s}}}$ and $J_{\rm{s}}^*$ if initial states of protocol (\ref{2}) are nonzero, and the assumption that initial states of protocol (\ref{2}) are zero is reasonable for practical multiagent networks.

In the proof of Theorem 1, the changing variable method is used to determine gain matrices ${K_u}$ and ${K_\phi }$, which makes the guaranteed-cost synchronization design criterion contain the nonlinear constraint ${P_x}{\hat P_x} = {I_n}$. However, if ${K_u}$ and ${K_\phi }$ are given previously, then this nonlinear constraint can be eliminated. The following corollary gives a leaderless guaranteed-cost synchronization analysis criterion.

{\emph{Corollary 1:}}~~For any given $J_{\rm{s}}^* > 0$, ${K_u}$ and ${K_\phi }$, multiagent network (\ref{1}) with protocol (\ref{2}) achieves leaderless guaranteed-cost synchronization if there exist $P_x^T = {P_x} > 0$ and $P_\phi ^T = {P_\phi } > 0$ such that
\[
{\hat \Theta _1} = {x^T}(0)\left( {\left( {{I_N} - {N^{ - 1}}{\bf{1}}{{\bf{1}}^T}} \right) \otimes {I_n}} \right)x(0){P_x} - J_{\rm{s}}^*{I_n} \le 0,
\]
\[
{\hat \Theta _j} = \left[ {\begin{array}{*{20}{c}}
   {{\Theta _{11}}} & {{\lambda _j}{P_\phi }{K_\phi }C} & {K_u^TR}  \\
   {*} & {\Theta _{22}^j} & {\bf{0}}  \\
   {*} & {\bf{0}} & { - R}  \\
\end{array}} \right] < 0{\rm{ }}\left( {j = 2,N} \right),\]
where ${\Theta _{11}} = {P_\phi }\left( {A + B{K_u}} \right) + {\left( {A + B{K_u}} \right)^T}{P_\phi }$ and $\Theta _{22}^j = {P_x}\left( {A + {\lambda _j}{K_\phi }C} \right) + $ ${\left( {A + {\lambda _j}{K_\phi }C} \right)^T}{P_x} + 2{\lambda _j}Q$.

In Theorem 1, the leaderless guaranteed-cost synchronization criterion contains a nonlinear constraint, which cannot be directly checked by LMI tools. Based on Corollary 1, the cone complementarity approach proposed by Ghaoui {\it et al.} in  \cite{GhaouiLE1997TAC} can deal with this nonlinear constraint by minimizing the trace of ${P_x}{\hat P_x}$. The feasibility problem of matrix inequalities in Theorem 1 can be transformed into the following minimization one:
\[
\begin{array}{l}
 {\rm{min {\kern 38pt} tr}}({P_x}{{\hat P}_x}) \\
 {\rm{sbuject~to {\kern 12pt}}}{{\hat \Xi }_1} < 0,{{\hat \Xi }_j} < 0{\rm{ }}(j = 2,N),{\rm{ }} \\
 {\rm{{\kern 56pt}}}{{\hat \Xi }_3} = \left[ {\begin{array}{*{20}{c}}
   {{P_x}} & I  \\
   {*} & {{{\hat P}_x}}  \\
\end{array}} \right] \ge 0. \\
 \end{array}\]
The following algorithm is presented to solve the above minimization problem.

\noindent\rule[0.25\baselineskip]{252pt}{1pt} \\
{\bf{Algorithm 1:}}

\noindent\rule[0.15\baselineskip]{252pt}{1pt}

{\bf{Step 1:}}~~Set $k = 0$. Check the feasibility of ${\hat \Xi _1} < 0,{\rm{ }}{\hat \Xi _j} < 0{\rm{ }}~(j = 2,N),$ and ${\hat \Xi _3} \ge 0$, and give ${P_{x,0}} = {P_x}$ and ${\hat P_{x,0}} = {\hat P_x}$.

{\bf{Step 2:}}~~Minimize the trace of ${P_x}{\hat P_{x,k}} + {P_{x,k}}{\hat P_x}$ subject to ${\hat \Xi _1} < 0,{\rm{ }}{\hat \Xi _j} < 0{\rm{ }}~(j = 2,N),$ and ${\hat \Xi _3} \ge 0$. Let ${P_{x,k + 1}} = {P_x}$ and ${\hat P_{x,k + 1}} = {\hat P_x}$.

{\bf{Step 3:}}~~Let ${K_u} = {\hat K_u}\hat P_\phi ^{ - 1}$ and ${K_\phi } =  - {\hat P_x}{C^T}$. If ${\hat \Theta _1} < 0$ and ${\rm{ }}{\hat \Theta _j} < 0{\rm{ }}~(j = 2,N)$ in Corollary 1 are feasible and $\left| {{\rm{tr(}}{P_x}{{\hat P}_x}) - 4n} \right| < \delta $ for some sufficiently small scalar $\delta  > 0$, then stop and give ${K_u}$ and ${K_\phi }$.

{\bf{Step 4:}}~~If $k$ is larger than the maximum allowed iteration number, then stop.

{\bf{Step 5:}}~~Set $ k = k+1 $ and go to Step 2.

\noindent\rule[0.25\baselineskip]{252pt}{1pt}

By the above analysis, ${{{{\hat x}_1}(t)} \mathord{\left/
 {\vphantom {{{{\hat x}_1}(t)} {\sqrt N }}} \right.
 \kern-\nulldelimiterspace} {\sqrt N }}$ is a valid candidate of the synchronization function. Due to ${\phi _j}(0) = {\bf{0}}{\rm{ }}\left( {j = 1,2, \cdots ,N} \right)$, one can obtain that ${\hat \phi _1}(t) \equiv {\bf{0}}$ and ${\dot {\hat x}}_1(t) = A{\hat x_1}(t)$ by (\ref{6}), which means that protocol states do not influence the synchronization function when initial protocol states are equal to zero. Moreover, it can be shown that ${\hat x_1}(0) = \left( {e_1^T{U^T} \otimes {I_n}} \right)x(0) = \sum\nolimits_{j = 1}^N {{{{x_j}(0)} \mathord{\left/
 {\vphantom {{{x_j}(0)} {\sqrt N }}} \right.
 \kern-\nulldelimiterspace} {\sqrt N }}} $, so the following corollary can be obtained, which gives an explicit expression of the synchronization function.

 {\emph{Corollary 2:}}~~If multiagent network (\ref{1}) with protocol (\ref{2}) achieves leaderless guaranteed-cost synchronization, then the synchronization function satisfies that
\[
\mathop {\lim }\limits_{t \to \infty } \left( {c(t) - \frac{1}{N}{{\rm e}^{At}}\sum\limits_{j = 1}^N {{x_j}(0)} } \right) = {\bf{0}}.\]

{\emph{Remark 4:}} Xiao {\it et al.} in \cite{XiaoF2007IET} first introduced the concept of the synchronization function to describe the whole feature of a multiagent network, where the synchronization protocol was constructed by state information of neighboring agents. For dynamic output feedback synchronization protocols, by Corollary 2, protocol states do not impact the synchronization function. Actually, if initial protocol states are not zero, then protocol states influence the explicit expression of the synchronization function in an input control way, which was shown in \cite{XiJ2012Auto}. Furthermore, the synchronization function is closely related to the autonomous dynamics of each agent and the average of initial states of all agents, and is identical for multiagent network (\ref{1}) with different undirected interaction topologies; that is, connected undirected interaction topologies with different structures do not impact the whole feature of multiagent networks. However, it should be also pointed out that this conclusion is no longer valid if the interaction topology is directed.

\section{Extensions to leader-following multiagent networks}\label{section4}
For high-order linear multiagent networks with leader-following structures and given cost budgets, this section gives guaranteed-cost synchronization design and analysis criteria, respectively, which are similar to leaderless cases, but the relationship between the cost budget and the LMI variable is different with leaderless cases.

For the leaderless multiagent networks, without loss of generality, we set that agent 1 is the leader and the other $N-1$ agents are followers. The whole interaction topology has a spanning tree with the root node representing the leader, where the leader does not receive any information from followers, only some followers can receive the outputs of the leader, and the local interaction topology among followers is undirected and can be unconnected. If multiagent network (\ref{1}) with a leader-following interaction topology achieves guaranteed-cost synchronization, then the synchronization function is the state of the leader; that is, ${\lim _{t \to \infty }}\left( {{x_j}(t) - {x_1}(t)} \right)$$ = {\bf{0}}{\rm{ }}\left( {j = 2,3, \cdots ,N} \right)$.

Since the leader does not receive any information and ${\phi _1}(0){\rm{ = }}{\bf{0}}$, one can obtain that ${\phi _1}(t){\rm{ = }}{\bf{0}}$. Hence, one has ${u_1}(t) \equiv {\bf{0}}$. Let ${\tilde x_j}(t) = {x_j}(t) - {x_1}(t){\rm{ }}\left( {j = 2,3, \cdots ,N} \right),$ $\tilde x(t) = {\left[ {\tilde x_2^T(t),\tilde x_3^T(t), \cdots ,\tilde x_N^T(t)} \right]^T}$, and $\tilde \phi (t) = \left[ {\phi _2^T(t),} \right. $ $ {\left. {\phi _3^T(t), \cdots ,\phi _N^T(t)} \right]^T}$, then the dynamics of multiagent network (\ref{1}) with protocol (\ref{2}) can be written as
\begin{eqnarray}\label{23}
\left\{ \begin{array}{l}
 \dot{ \tilde x}(t) = \left( {{I_{N - 1}} \otimes A} \right)\tilde x(t) + \left( {{I_{N - 1}} \otimes B{K_u}} \right)\tilde \phi (t), \\
 \dot {\tilde \phi}(t) = \left( {{I_{N - 1}} \otimes (A + B{K_u}) + \left( {{L_{ff}} + {\Lambda _{fl}}} \right)} \right. \\
 {\kern 26pt} \left. { \otimes {K_\phi }C} \right)\tilde \phi (t) - \left( {\left( {{L_{ff}} + {\Lambda _{fl}}} \right) \otimes {K_\phi }C} \right)\tilde x(t), \\
 \end{array} \right.
\end{eqnarray}
where ${L_{ff}}$ is the Laplacian matrix of the interaction topology among followers and ${\Lambda _{fl}} = {\rm{diag}}\left\{ {{w_{21}},{w_{31}}, \cdots ,{w_{N1}}} \right\}$ denotes the interaction from the leader to followers. Let ${l_{fl}} = {\left[ {{w_{21}},{w_{31}}, \cdots ,{w_{N1}}} \right]^T}$, then the Laplacian matrix of the whole interaction topology is
\[
L = \left[ {\begin{array}{*{20}{c}}
   0 & {\bf{0}}  \\
   { - {l_{fl}}} & {{L_{ff}} + {\Lambda _{fl}}}  \\
\end{array}} \right].
\]
Since the whole interaction topology has a spanning tree and the local interaction topology among followers is undirected, there exists an orthonormal matrix $\tilde U$ such that ${\tilde U^T}\left( {{L_{ff}} + {\Lambda _{fl}}} \right)\tilde U = {\rm{diag}}\left\{ {{\lambda _2},{\lambda _3}, \cdots ,{\lambda _N}} \right\}$ with $0 < {\lambda _2} \le {\lambda _3} \le  \cdots  \le {\lambda _N}$ being nonzero eigenvalues of $L$. Let
\[
\left( {{{\tilde U}^T} \otimes {I_n}} \right)\tilde x(t) = {\left[ {\mathord{\buildrel{\lower3pt\hbox{$\scriptscriptstyle\frown$}}
\over x} _2^T(t),\mathord{\buildrel{\lower3pt\hbox{$\scriptscriptstyle\frown$}}
\over x} _3^T(t), \cdots ,\mathord{\buildrel{\lower3pt\hbox{$\scriptscriptstyle\frown$}}
\over x} _N^T(t)} \right]^T},
\]
\[
\left( {{{\tilde U}^T} \otimes {I_n}} \right)\tilde \phi (t) = {\left[ {\mathord{\buildrel{\lower3pt\hbox{$\scriptscriptstyle\frown$}}
\over \phi } _2^T(t),\mathord{\buildrel{\lower3pt\hbox{$\scriptscriptstyle\frown$}}
\over \phi } _3^T(t), \cdots ,\mathord{\buildrel{\lower3pt\hbox{$\scriptscriptstyle\frown$}}
\over \phi } _N^T(t)} \right]^T},
\]
then one can obtain by (\ref{23}) that
\begin{eqnarray}\label{24}
\left\{ \begin{array}{l}
{\kern -6pt}  {{\dot {\mathord{\buildrel{\lower3pt\hbox{$\scriptscriptstyle\frown$}}
\over x} }}_j}(t) {\kern -2pt} = {\kern -2pt} A{{\mathord{\buildrel{\lower3pt\hbox{$\scriptscriptstyle\frown$}}
\over x} }_j}(t) + B{K_u}{{\mathord{\buildrel{\lower3pt\hbox{$\scriptscriptstyle\frown$}}
\over \phi } }_j}(t), \\
{\kern -6pt} {{\dot {\mathord{\buildrel{\lower3pt\hbox{$\scriptscriptstyle\frown$}}
\over \phi } }}_j}(t) {\kern -2pt} = {\kern -2pt} \left( {A + B{K_u} + {\lambda _j}{K_\phi }C} \right){{\mathord{\buildrel{\lower3pt\hbox{$\scriptscriptstyle\frown$}}
\over \phi } }_j}(t) {\kern -2pt}-{\kern -2pt} {\lambda _j}{K_\phi }C{{\mathord{\buildrel{\lower3pt\hbox{$\scriptscriptstyle\frown$}}
\over x} }_j}(t), \\
 \end{array} \right.
\end{eqnarray}
where $j = 2,3, \cdots ,N.$ One can find that if ${\lim _{t \to \infty }}\tilde x(t) = {\bf{0}}$, then multiagent network (\ref{1}) with protocol (\ref{2}) achieves leader-following synchronization, which is equivalent to ${\lim _{t \to \infty }}{\mathord{\buildrel{\lower3pt\hbox{$\scriptscriptstyle\frown$}}
\over x} _j}(t) = {\bf{0}}$ $ \left( {j = 2,3, \cdots ,N} \right)$
since $\tilde U \otimes {I_n}$ is nonsingular. For the symmetric and positive ${P_x}$, it can be shown that
\begin{eqnarray}\label{25}
\begin{array}{l}
 \sum\limits_{j = 2}^N {\mathord{\buildrel{\lower3pt\hbox{$\scriptscriptstyle\frown$}}
\over x} _j^T(0){P_x}{{\mathord{\buildrel{\lower3pt\hbox{$\scriptscriptstyle\frown$}}
\over x} }_j}(0)}  \\
 {\kern 28pt} = {x^T}(0)\left( {\left[ {\begin{array}{*{20}{c}}
   {N - 1} & { - {\bf{1}}_{N - 1}^T}  \\
   { - {{\bf{1}}_{N - 1}}} & {{I_{N - 1}}}  \\
\end{array}} \right] \otimes {P_x}} \right)x(0). \\
 \end{array}
\end{eqnarray}

Based on the above facts, by the similar analysis to Theorem 1 and Corollary 1, sufficient conditions for leader-following guaranteed-cost synchronization design and analysis with the given cost budget are given as follows.

{\emph{Theorem 2:}}~~For any given $J_{\rm{s}}^* > 0$, multiagent network (\ref{1}) is leader-following guaranteed-cost synchronizable by protocol (\ref{2}) if there exist $P_x^T = {P_x} > 0,$ $\hat P_x^T = {\hat P_x} > 0,$ $\hat P_\phi ^T = {\hat P_\phi } > 0,$ and ${\hat K_u }$ such that
\[
{x^T}(0)\left( {\left[ {\begin{array}{*{20}{c}}
   {N - 1} & { - {\bf{1}}_{N - 1}^T}  \\
   { - {{\bf{1}}_{N - 1}}} & {{I_{N - 1}}}  \\
\end{array}} \right] \otimes {I_n}} \right)x(0){P_x} - J_{\rm{s}}^*{I_n} \le 0,\]
\[
\left[ {\begin{array}{*{20}{c}}
   {{\Xi _{11}}} & { - {\lambda _j}{{\hat P}_x}{C^T}C} & {\hat K_u^TR}  \\
   {*} & {\Xi _{22}^j} & {\bf{0}}  \\
   {*} & {\bf{0}} & { - R}  \\
\end{array}} \right] < 0\left( {j = 2,N} \right),\]
\[
{P_x}{\hat P_x} = {I_n},\]
where ${\Xi _{11}} = A{\hat P_\phi } + {\hat P_\phi }{A^T} + B{\hat K_u} + \hat K_u^T{B^T}$ and $\Xi _{22}^j = {P_x}A + {A^T}{P_x} - 2{\lambda _j}{C^T}C + 2{\lambda _j}Q.$ In this case, ${K_u} = {\hat K_u}\hat P_\phi ^{ - 1}$ and ${K_\phi } =  - {\hat P_x}{C^T}$.

{\emph{Corollary 3:}}~~For any given $J_{\rm{s}}^* > 0$, ${K_u}$ and ${K_\phi }$, multiagent network (\ref{1}) with protocol (\ref{2}) achieves leader-following guaranteed-cost synchronization if there exist $P_x^T = {P_x} > 0,$ $P_\phi ^T = {P_\phi } > 0$ such that
\[
{x^T}(0)\left( {\left[ {\begin{array}{*{20}{c}}
   {N - 1} & { - {\bf{1}}_{N - 1}^T}  \\
   { - {{\bf{1}}_{N - 1}}} & {{I_{N - 1}}}  \\
\end{array}} \right] \otimes {I_n}} \right)x(0){P_x} - J_{\rm{s}}^*{I_n} \le 0,
\]
\[
\left[ {\begin{array}{*{20}{c}}
   {{\Theta _{11}}} & {{\lambda _j}{P_\phi }{K_\phi }C} & {K_u^TR}  \\
   {*} & {\Theta _{22}^j} & {\bf{0}}  \\
   {*} & {\bf{0}} & { - R}  \\
\end{array}} \right] < 0{\rm{ }}\left( {j = 2,N} \right),\]
where ${\Theta _{11}} = {P_\phi }\left( {A + B{K_u}} \right) + {\left( {A + B{K_u}} \right)^T}{P_\phi }$ and $\Theta _{22}^j = {P_x}\left( {A + {\lambda _j}{K_\phi }C} \right) + $ ${\left( {A + {\lambda _j}{K_\phi }C} \right)^T}{P_x} + 2{\lambda _j}Q$.

By the cone complementarity approach, the feasible problem of the matrix inequalities in Theorem 2 can also be converted into a minimization one, which can be checked by a similar algorithm to Algorithm 1. Here, the detail description is omitted due to the length limitation.

Furthermore, the variable changing method and the cone complementarity approach are applied to determine gain matrices of synchronization protocols. The variable changing method does not introduce any conservatism since it is an equivalent transformation. However, the cone complementarity approach may bring in some conservatism to deal with the impacts of nonlinearity. In \cite{GhaouiLE1997TAC}, the conservatism of the cone complementarity approach was discussed detailedly and it was shown that less conservatism may be introduced by numerical simulations.

Moreover, there are three key difficulties in obtaining Theorems 1 and 2. The first one is to construct the relationship between the linear quadratic optimization index and the Laplacian matrix of the interaction topology, as shown in (16) and (17). The second one is to construct the relationship between the given cost budget and the variable of LMI criteria, as given in (19) and (20). The third one is to transform the leader-following synchronization problem into the leaderless one with the different structure matrix, as shown in (23) and (24).

{\emph{Remark 5:}} For guaranteed-cost synchronization criteria of leaderless and leader-following multiagent networks, the key distinction is that the relationship matrices between the given cost budget and the LMI variable are different. For leaderless cases, the relationship matrix ${I_N} - {N^{ - 1}}{\bf{1}}{{\bf{1}}^T}$ is the Laplacian matrix of a complete graph with edge weights ${N^{ - 1}}$. For leader-following cases, the relationship matrix $\left[ {\begin{array}{*{20}{c}}
   {N - 1} & { - {\bf{1}}_{N - 1}^T}  \\
   { - {{\bf{1}}_{N - 1}}} & {{I_{N - 1}}}  \\
\end{array}} \right]$ is the Laplacian matrix of a star graph with edge weights 1 and the central node is the leader. The two relationships intrinsically reflect the structure characteristics of multiagent networks; that is, the average of the initial states of all agents determines the whole motion for leaderless structures, but the whole motion only depends on the leader for leader-following structures.

{\emph{Remark 6:}} The LMI criteria for guaranteed-cost synchronization are dependent on the Laplacian matrices of interaction topologies in \cite{ZhaoY2015ISA}-\hspace{-0.5pt}\cite{Zhao2017ISA}. In this case, the dimensions of the variables are identical with the number of agents, so it is time-cost to check those criteria when multiagent networks consist of a large number of agents. However, the LMI criteria in Theorems 1 and 2 are only dependent on the minimum and maximum nonzero eigenvalues of the Laplacian matrix, so the computational complexity is lower. Meanwhile, it should be pointed out that the cone complementarity approach is used to deal with the impacts of nonlinear terms in Theorems 1 and 2. Because this method is an iteration algorithm, the computational complexity may increase and the associated algorithm may be not robust. Ghaoui {\it et al.} in \cite{GhaouiLE1997TAC} showed that this method is robust and has lower computational complexity by many numerical simulations.

\section{Illustrative examples}\label{section5}
In this section, two numerical examples are presented to illustrate the effectiveness of main results on leaderless and leader-following multiagent networks, respectively.

A 3-dimensional multiagent network is considered, where it is composed of six agents labeled from 1 to 6. The dynamics of each agent is described as (\ref{1}) with
\[
A = \left[ {\begin{array}{*{20}{c}}
   {0.2} & {3.5} & 0  \\
   { - 1.5} & {0.8} & { - 1.3}  \\
   1 & 0 & { - 2.6}  \\
\end{array}} \right],
\]
\[
B = \left[ {\begin{array}{*{20}{c}}
   2 & 0  \\
   { - 1.5} & 4  \\
   0 & { - 0.4}  \\
\end{array}} \right],
\]
\[
{C = \left[ {\begin{array}{*{20}{c}}
   2 & 0 & 2  \\
   { - 1.5} & 3 & 0  \\
\end{array}} \right].}\]
The initial states are
\[
\begin{array}{l}
 {x_1}(0) = {[ - {\rm{13, 20, }} - {\rm{3}}]^T}, {\kern 5pt}{x_2}(0) = {[ - {\rm{16, }} - {\rm{8,  15}}]^T},{\rm{ }} \\
 {x_3}(0) = {[{\rm{26, 10,}} - {\rm{12}}]^T}, {\kern 8pt} {x_4}(0) = {[ - {\rm{3,}} - {\rm{8, 19}}]^T},{\rm{   }} \\
 {x_5}(0) = {[{\rm{12, 22,}} - {\rm{6}}]^T},{\kern 13pt}{x_6}(0) = {[{\rm{8,}} - {\rm{13, 16}}]^T}. \\
 \end{array}\]
The interaction topologies for {\it the Leaderless case} and {\it the Leader-following case} are respectively given as $G_1$ and $G_2$ in Fig. 1.

\begin{figure}[!htb]
\begin{center}
\scalebox{0.8}[0.8]{\includegraphics{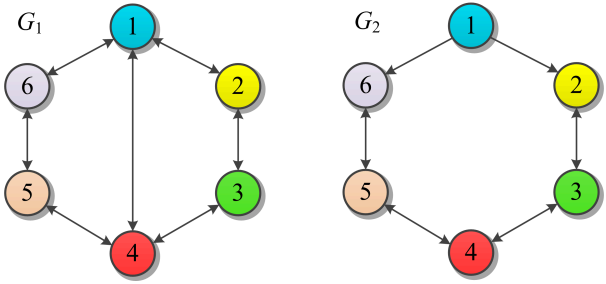}}\vspace{0pt}
\caption{The interaction topology $G$.}\vspace{0pt}
\end{center}
\end{figure}

{\bf{Example 1:}}~~({\it Leaderless case}) The interaction topology $G_1$ is given as Fig.1, where the weights of edges of the interaction topology are 1. In the linear quadratic optimization index, the matrices $Q$ and $R$ are given as
\[
Q = \left[ {\begin{array}{*{20}{c}}
   {0.3} & {0.06} & 0  \\
   {0.06} & {0.3} & {0.06}  \\
   0 & {0.06} & {0.3}  \\
\end{array}} \right],
\]
\[
R = \left[ {\begin{array}{*{20}{c}}
   {0.8} & {0.08}  \\
   {0.08} & {0.8}  \\
\end{array}} \right].
\]
The given cost budget is $J_{\rm{s}}^* = 6000$, which is an upper bound of the linear quadratic index in (\ref{2}) and includes the energy consumption and the synchronization regulation performance. Thus, according to Algorithm 1, one has
\[
{K_u} = \left[ {\begin{array}{*{20}{c}}
   {5.1141} & {8.0251} & { - 0.5324}  \\
   { - 44.4484} & { - 63.8269} & {3.1964}  \\
\end{array}} \right],\]
\[
{K_\phi } = \left[ {\begin{array}{*{20}{c}}
   { - 2.1446} & {1.1269}  \\
   { - 0.3219} & { - 1.6096}  \\
   { - 1.1376} & { - 0.0013}  \\
\end{array}} \right].\]
It should be pointed out that ${K_u}$ and ${K_\phi}$ cannot be determined by Algorithm 1 if the limited cost budget cannot provide the enough energy. In this case, ${\hat \Xi _1} \le 0$ in Theorem 1 is not feasible.

The state trajectories of the multiagent network are shown in Figs. 2 to 4, where the trajectories marked by circles denote the curves of the synchronization function $c(t)$ obtained by Corollary 2, which satisfies ${\lim _{t \to \infty }}(c(t) - {{\rm e} ^{At}} {[2.3333,3.8333,4.8333]^T}) = {\bf 0}$. Fig. 5 shows the trajectories of the linear quadratic optimization index. It is clear that this multiagent network achieves leaderless guaranteed-cost synchronization with the given cost budget.
 

\begin{figure}[!htb]
\begin{center}
\scalebox{0.46}[0.46]{\includegraphics{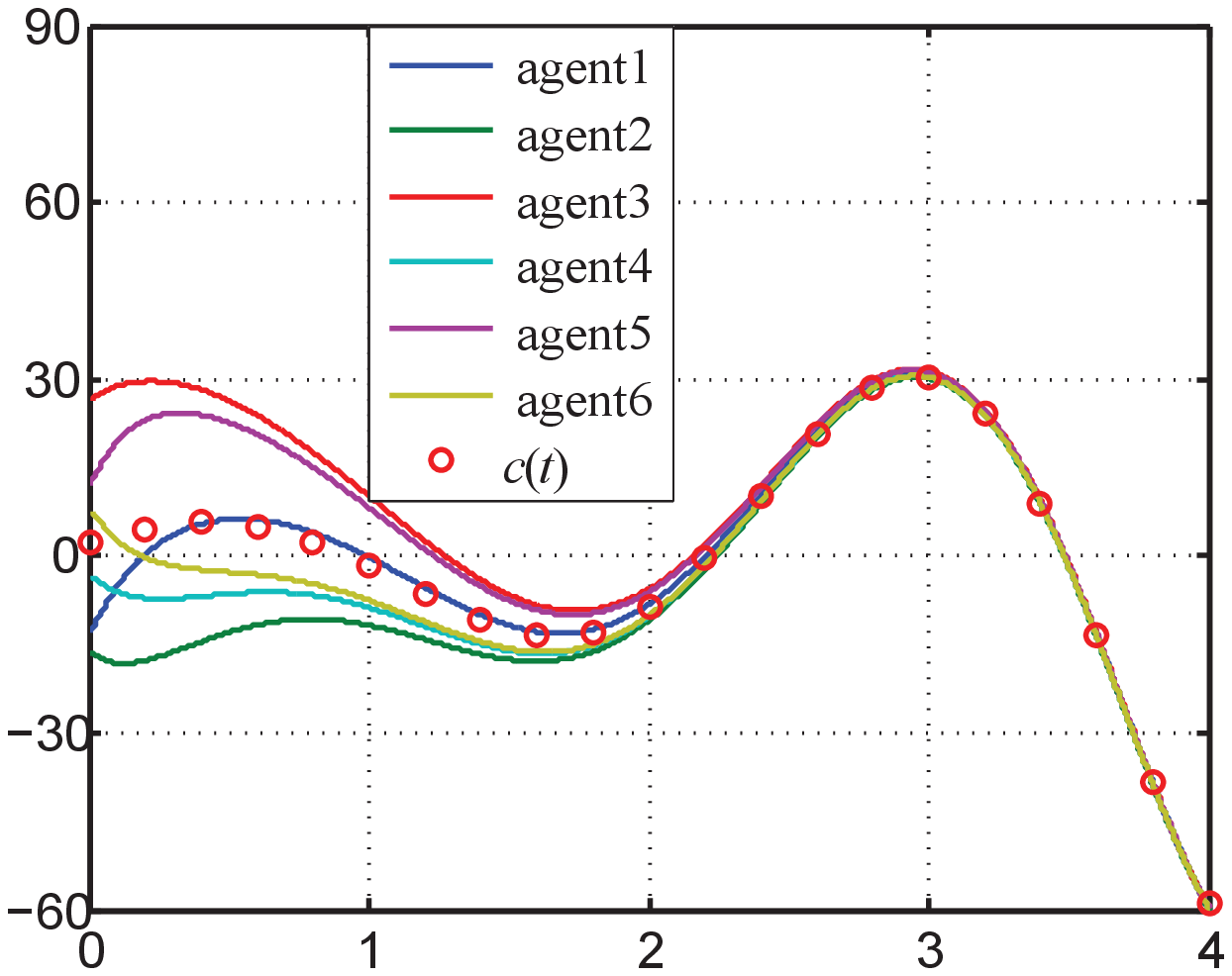}}\vspace{-5pt}
\put (-190, 36) {\rotatebox{90} {{\scriptsize ${x_{j1}}(t)~(j = 1,2, \cdots ,6)$}}}
\put (-97, -5) {{\scriptsize Time}}\vspace{5pt}
\caption{State trajectories of ${x_{j1}}(t)~(j = 1,2, \cdots ,6)$.}
\end{center}\vspace{0pt}
\end{figure}
\vspace{0pt}

\begin{figure}[!htb]
\begin{center}
\scalebox{0.46}[0.46]{\includegraphics{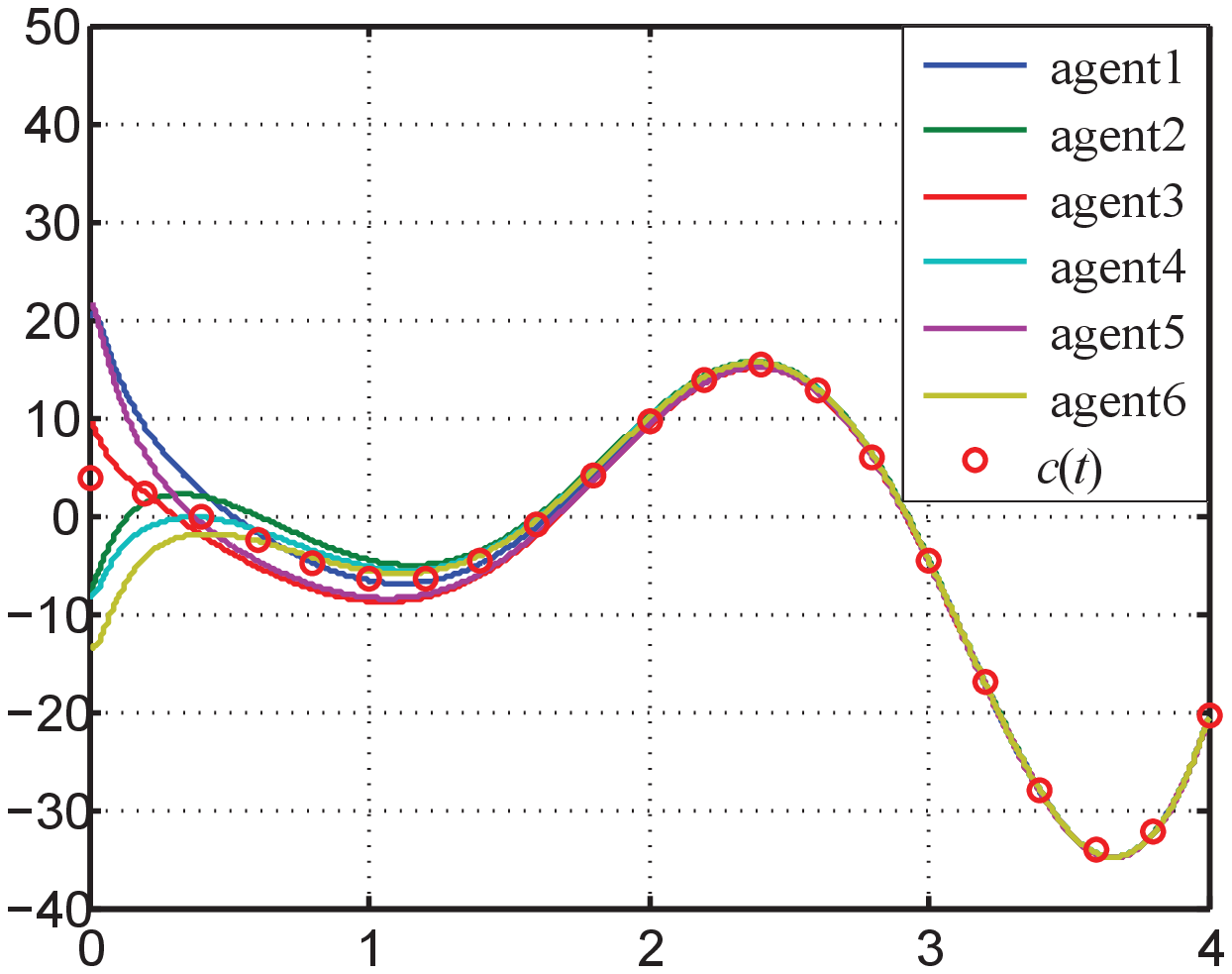}}\vspace{-5pt}
\put (-190, 36)  {\rotatebox{90} {{\scriptsize ${x_{j2}}(t)~(j = 1,2, \cdots ,6)$}}}
\put (-97, -5) {{\scriptsize Time}}\vspace{5pt}
\caption{State trajectories of ${x_{j2}}(t)~(j = 1,2, \cdots ,6)$.}
\end{center}\vspace{0pt}
\end{figure}
\vspace{0pt}

\begin{figure}[!htb]
\begin{center}
\scalebox{0.46}[0.46]{\includegraphics{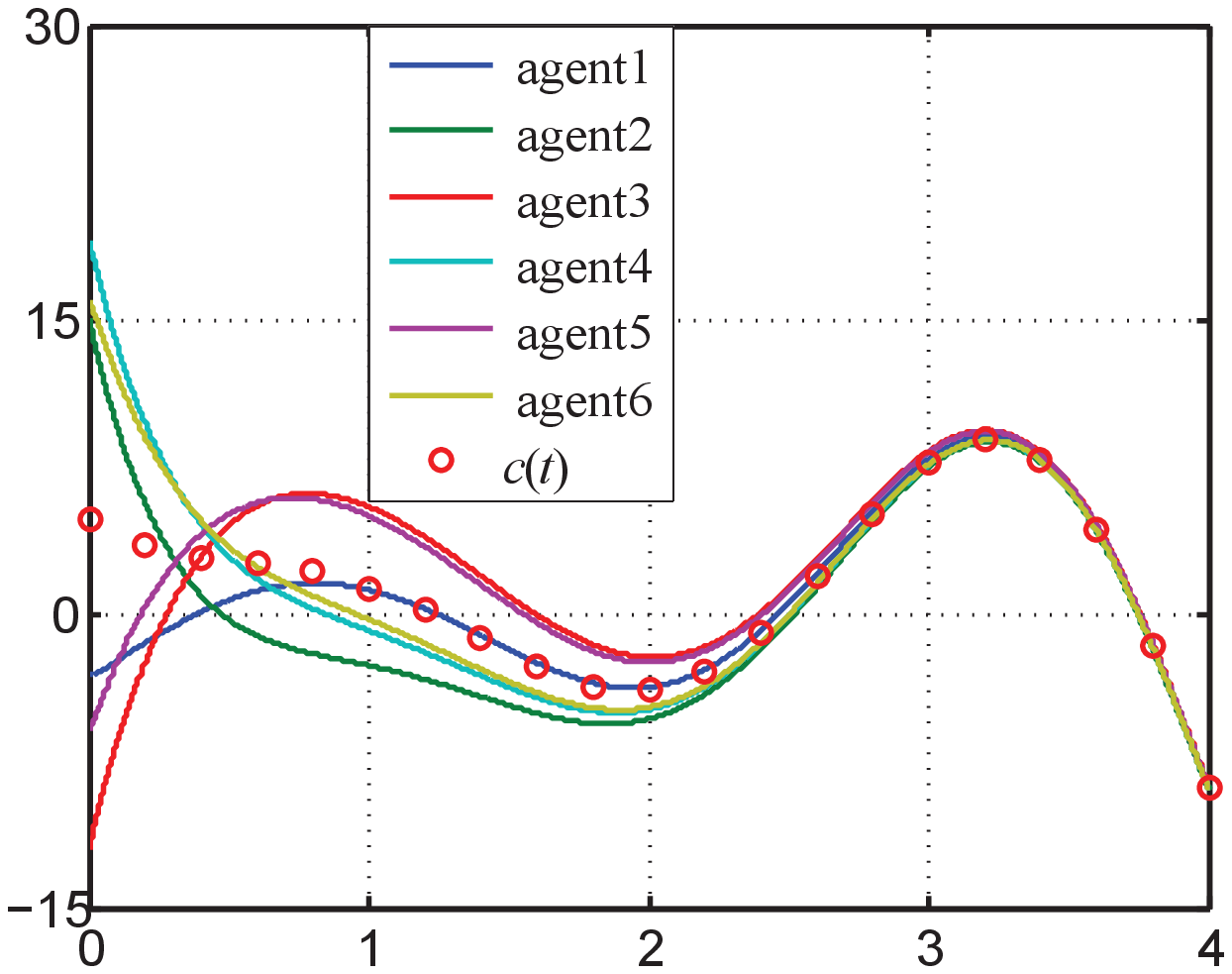}} \vspace{-5pt}
\put (-190, 36)  {\rotatebox{90} {{\scriptsize ${x_{j3}}(t)~(j = 1,2, \cdots ,6)$}}}
\put (-100, -5) {{\scriptsize Time}}\vspace{5pt}
\caption{State trajectories of ${x_{j3}}(t)~(j = 1,2, \cdots ,6)$.}
\end{center}\vspace{0pt}
\end{figure}
\vspace{0pt}

\begin{figure}[!htb]
\begin{center}
\scalebox{0.46}[0.46]{\includegraphics{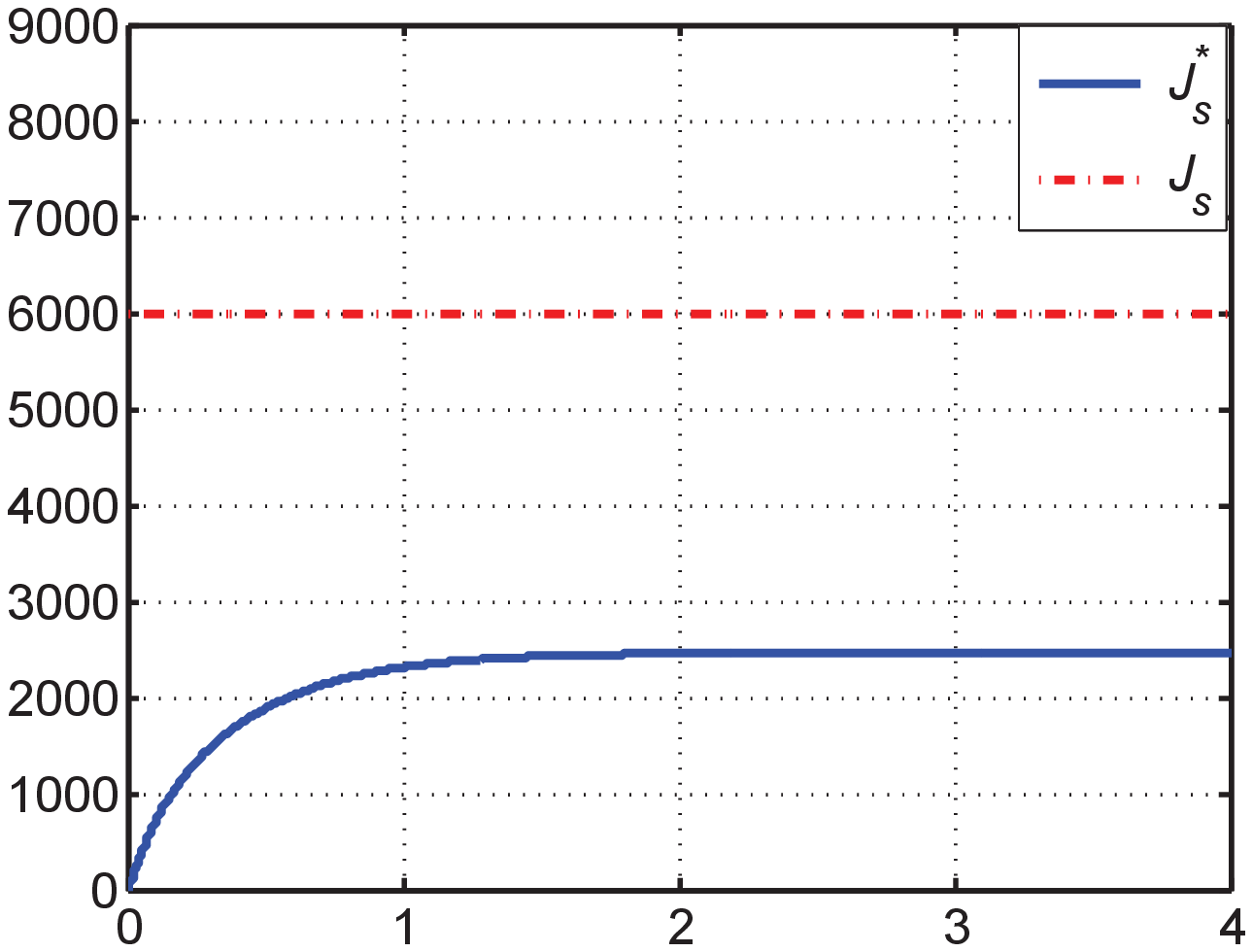}} \vspace{-5pt}
\put (-200, 66)  {\rotatebox{90} {{\scriptsize Cost}}}
\put (-103, -5) {{\scriptsize Time}}\vspace{5pt}
\caption{Trajectories of cost.}
\end{center}\vspace{-0.45em}
\end{figure}
\vspace{0pt}

%
%

{\bf{Example 2:}}~~({\it Leader-following case}) In this case, agent 1 is the leader and the other 5 agents are followers. The interaction topology $G_2$ is given as Fig. 1, where the weights of edges are 1. In this case, it is set that

\[
Q = \left[ {\begin{array}{*{20}{c}}
   {0.25} & {0.05} & {0.05}  \\
   {0.05} & {0.25} & 0  \\
   {0.05} & 0 & {0.25}  \\
\end{array}} \right],
\]
\[
R = \left[ {\begin{array}{*{20}{c}}
   {0.75} & {0.15}  \\
   {0.15} & {0.75}  \\
\end{array}} \right].
\]
The given cost budget is $J_{\rm{s}}^* = 10000$. Thus, according to Theorem 2, one has
\[
{K_u} = \left[ {\begin{array}{*{20}{c}}
   {{\rm{1}}{\rm{.5586}}} & {{\rm{2}}{\rm{.8988}}} & {{\rm{ - 0}}{\rm{.271}}7}  \\
   {{\rm{ - 8}}{\rm{.896}}6} & {{\rm{ - 12}}{\rm{.9081}}} & {{\rm{0}}{\rm{.6719}}}  \\
\end{array}} \right],\]
\[{K_\phi } = \left[ {\begin{array}{*{20}{c}}
   {{\rm{ - 2}}{\rm{.5652}}} & {{\rm{1}}{\rm{.6794}}}  \\
   {{\rm{ - 0}}{\rm{.3525}}} & {{\rm{ - 1}}{\rm{.9704}}}  \\
   {{\rm{ - 1}}{\rm{.0104}}} & {{\rm{ - 0}}{\rm{.284}}3}  \\
\end{array}} \right].
\]
The trajectories of state errors $ {x_j}(t) - {x_1}(t){\rm{ }}\left( {j = 2,3, \cdots ,N} \right)$ of this multiagent network are shown in Figs. 9 to 11, and the trajectories of the linear quadratic optimization index are given in Fig. 12. Thus, it can be found that this multiagent network achieves leader-following guaranteed-cost synchronization with the given cost budget. 

\vspace{1.045em}
\begin{figure}[!htb]
\begin{center}
\scalebox{0.46}[0.46]{\includegraphics{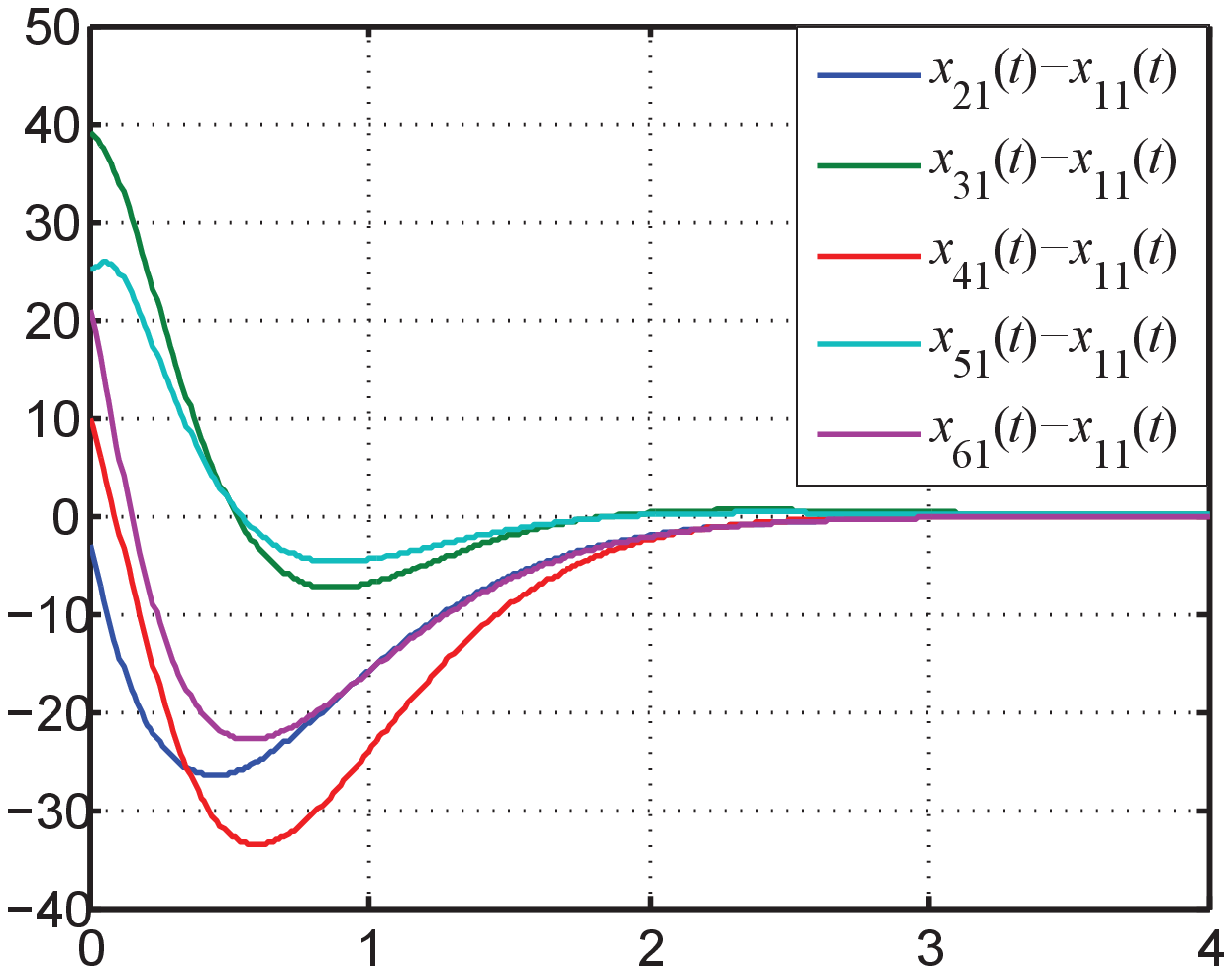}}\vspace{-5pt}
\put (-190, 26) {\rotatebox{90} {{\scriptsize ${x_{j1}}(t) - {x_{11}}(t) ~(j = 2, \cdots ,6)$}}}
\put (-97, -5)  {{\scriptsize Time}}\vspace{3pt}
\caption{State trajectories of ${x_{j1}}(t) - {x_{11}}(t) ~(j = 2, \cdots ,6)$.}
\end{center}\vspace{-9pt}
\end{figure}

\begin{figure}[!htb]
\begin{center}
\scalebox{0.46}[0.46]{\includegraphics{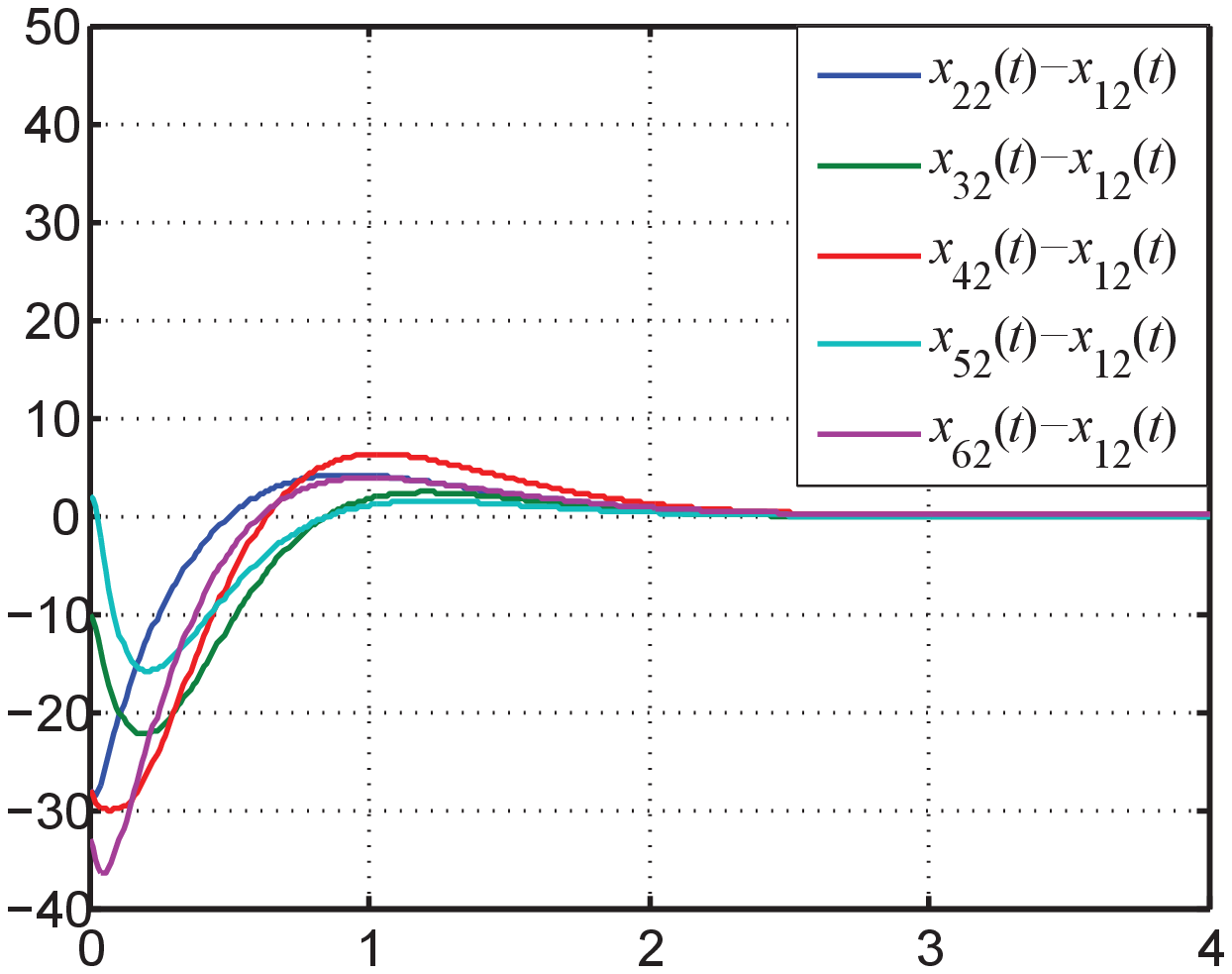}}\vspace{-5pt}
\put (-190, 26) {\rotatebox{90} {{\scriptsize ${x_{j2}}(t) - {x_{12}}(t) ~(j = 2, \cdots ,6)$}}}
\put (-97, -5)  {{\scriptsize Time}}\vspace{3pt}
\caption{State trajectories of ${x_{j2}}(t) - {x_{12}}(t) ~(j = 2, \cdots ,6)$.}
\end{center}\vspace{17pt}
\end{figure}
\vspace{0pt}

\begin{figure}[!htb]
\begin{center}
\scalebox{0.46}[0.46]{\includegraphics{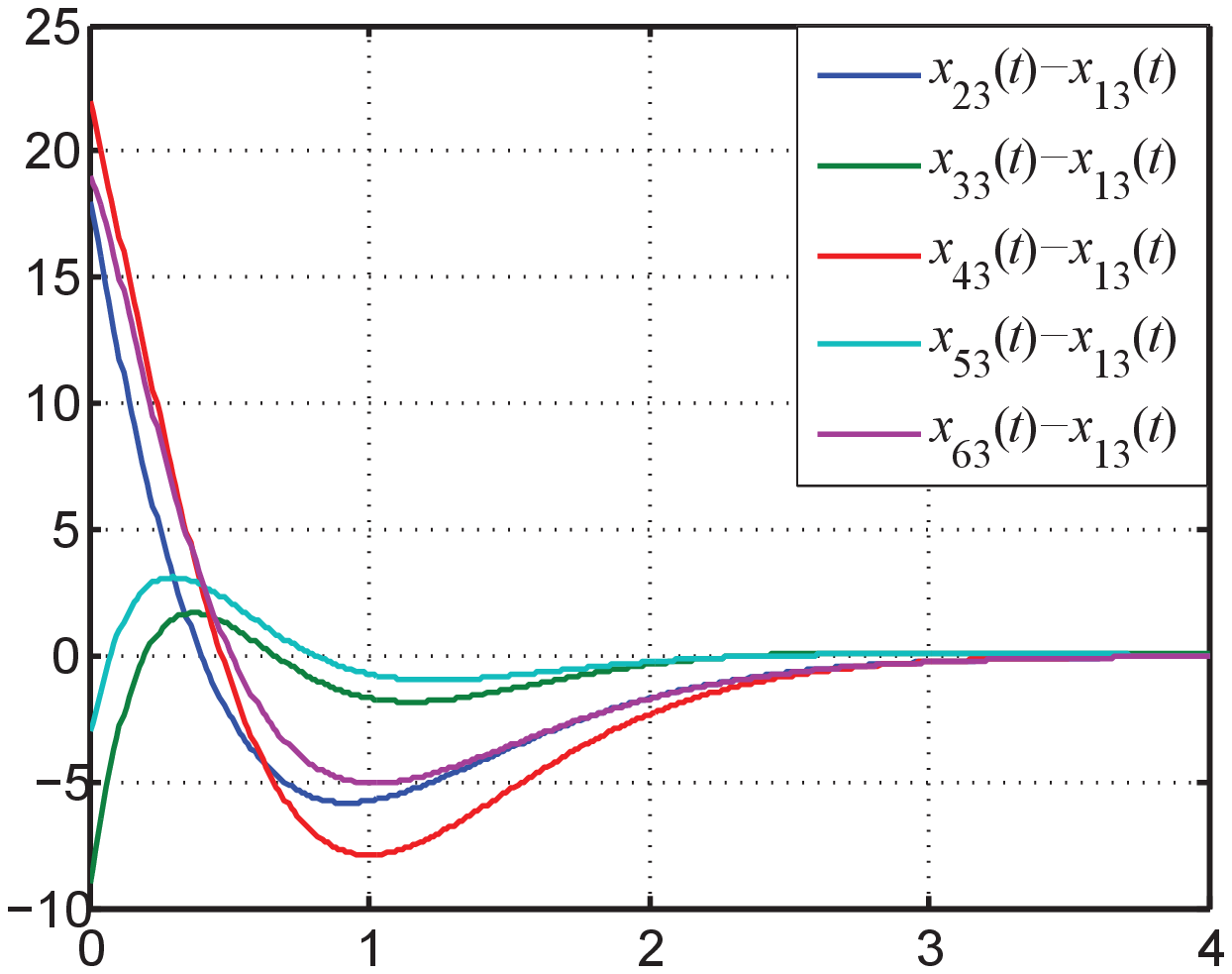}} \vspace{-5pt}
\put (-190, 26) {\rotatebox{90} {{\scriptsize ${x_{j3}}(t) - {x_{13}}(t) ~(j = 2, \cdots ,6)$}}}
\put (-100, -5)  {{\scriptsize Time}}\vspace{3pt}
\caption{State trajectories of ${x_{j3}}(t) - {x_{13}}(t) ~(j = 2, \cdots ,6)$.}
\end{center}\vspace{-7pt}
\end{figure}
\vspace{0pt}

\begin{figure}[!htb]
\begin{center}
\scalebox{0.46}[0.46]{\includegraphics{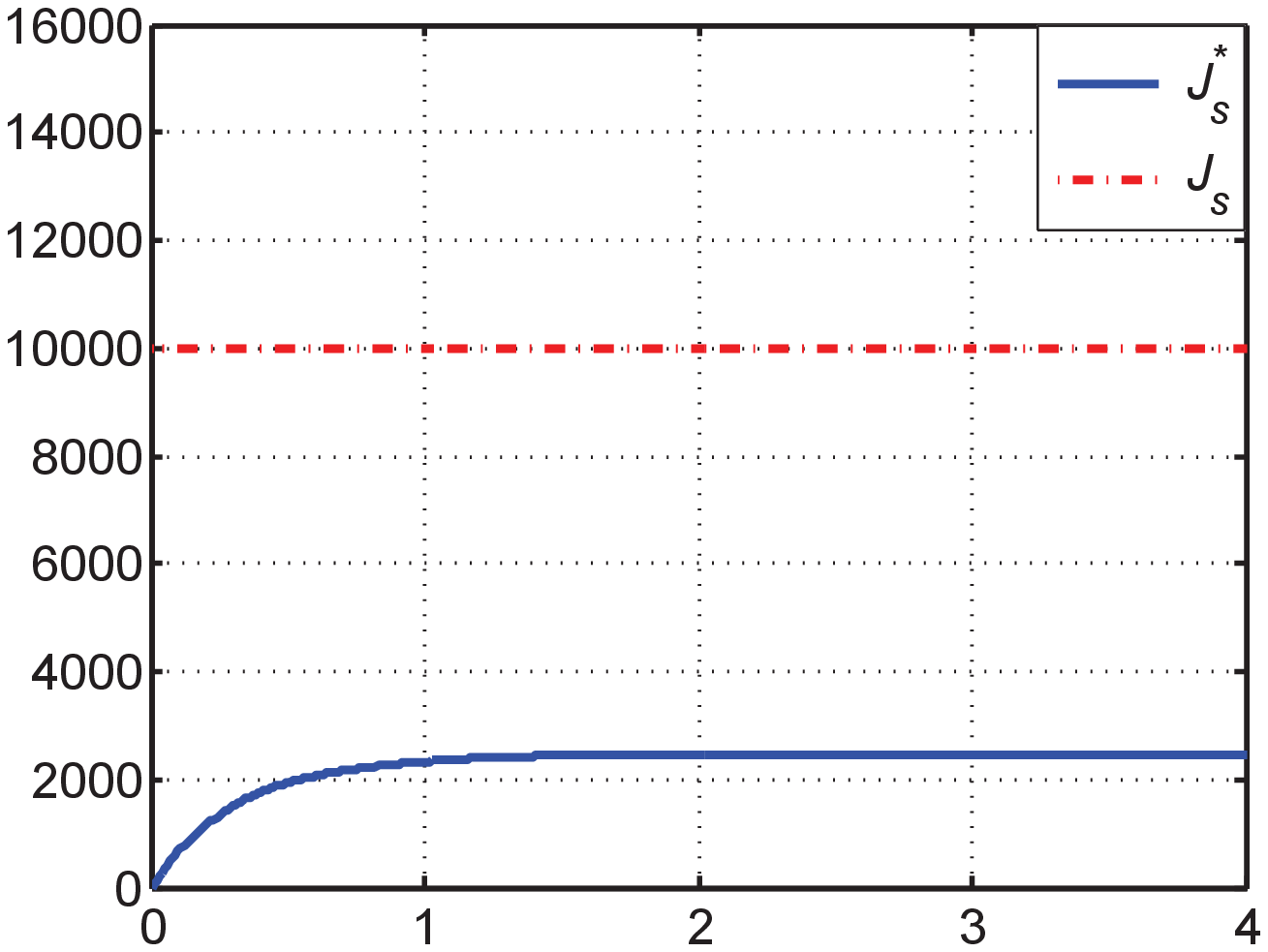}} \vspace{-5pt}
\put (-200, 66)  {\rotatebox{90} {{\scriptsize Cost}}}
\put (-103, -5)  {{\scriptsize Time}}\vspace{3pt}
\caption{Trajectories of cost.}
\end{center}\vspace{-7pt}
\end{figure}
\vspace{0pt}

\section{Conclusion}\label{section6}
Both leaderless and leader-following guaranteed-cost synchronization analysis and design problems for multiagent networks with the given cost budget were investigated by using output information of neighboring agents. The guaranteed-cost synchronization analysis and design criteria independent of the number of agents were proposed by constructing dynamic output feedback synchronization protocols and the relationships between the given cost budget and the LMI variable, where synchronization protocols satisfy a specific separation principle and those relationships depend on the structures of interaction topologies. Especially, the specific separation principle can simplify the synchronization design, but nonlinear terms are still introduced due to guaranteed-cost constraints. Moreover, an algorithm was presented to deal with nonlinear constraints and to determine gain matrices of synchronization protocols.

Furthermore, the future research topic can focus on two aspects. The first one is to deal with the impacts of time-varying delays and directed interaction topologies on guaranteed-cost synchronization of multiagent networks with dynamic output feedback synchronization protocols. The second one is to investigate the practical applications of multiagent networks combining the main results in the current paper with structure features of practical multiagent networks, such as multiple agent supporting systems, network congestion control systems and single-link manipulator systems with a flexible joint, {\it et al.} \par

\section{Acknowledgement}
The authors would like to thank Jie Gao for providing numerical analysis and simulation.

\ifCLASSOPTIONcaptionsoff
\newpage
\fi

\end{document}